\documentclass[twocolumn,tighten]{aastex631}
\hypersetup{linkcolor=red,citecolor=blue,filecolor=cyan,urlcolor=magenta}
\usepackage{amsmath}
\usepackage{natbib}
\usepackage{todonotes}
\usepackage{CJK}

\newcommand{\degree}{$^{\circ}$}

\newcommand{\zmax}{$Z_\text{max}$}
\newcommand{\vphi}{$v_\phi$}
\newcommand{\Astropy}{\textit{Astropy}}
\newcommand{\Gaia}{\textit{Gaia~}}
\newcommand{\Nstack}{$N_{\text{stack}}$}
\newcommand{\Nstackcut}{$N_{\text{stack}}^{\text{cut}}$}
\newcommand{\HDBSCAN}{\texttt{HDBSCAN}}
\newcommand{\RN}[1]{%
  \textup{\uppercase\expandafter{\romannumeral#1}}%
}

\shorttitle{Milky Way Substructure}
\shortauthors{Ou, Necib, \& Frebel}

\begin{document}
\begin{CJK*}{UTF8}{gbsn}

\title{Robust Clustering of the Local Milky Way Stellar Kinematic Substructures with \Gaia eDR3}

\author[0000-0002-4669-9967]{Xiaowei Ou (欧筱葳)} 
\affiliation{%
Kavli Institute for Astrophysics and Space Research,
Massachusetts Institute of Technology,
77 Massachusetts Ave, Cambridge MA 02139, USA}
\email{Email:\ xwou@mit.edu}

\author[0000-0003-2806-1414]{Lina Necib}
\affiliation{Kavli Institute for Astrophysics and Space Research,
Massachusetts Institute of Technology,
77 Massachusetts Ave, Cambridge MA 02139, USA}
\affiliation{The NSF AI Institute for Artificial Intelligence and Fundamental Interactions,
Massachusetts Institute of Technology,
77 Massachusetts Ave, Cambridge MA 02139, USA}

\author{Anna Frebel}
\affiliation{Kavli Institute for Astrophysics and Space Research,
Massachusetts Institute of Technology,
77 Massachusetts Ave, Cambridge MA 02139, USA}

\begin{abstract}

We apply the clustering algorithm \HDBSCAN~on the \Gaia early third data release astrometry combined with the \Gaia second data release radial velocity measurements of almost 5.5 million stars to identify the local stellar kinematic substructures in the solar neighborhood. 
Understanding these structures helps build a more complete picture of the formation of the Milky Way, as well as an empirical phase space distribution of dark matter that would inform detection experiments.
The main goal of this study is to provide a list of the most stable clusters, by taking into account the measurement uncertainties and studying the stability of the clustering results.
We apply the clustering algorithm in two spaces, in velocity space in order to study recently accreted structures, and in action-angle space to find phase-mixed structures.
We find 1405 (497) stars in 23 (6) robust clusters in velocity space (action-angle space) that are consistently not associated with noise. They are attributed to the known structures: the Gaia Sausage-Enceladus, the Helmi Stream, and globular cluster NGC~3201 are found in both spaces, while NGC~104 and the thick disk (Sequoia) are identified in velocity space (action-angle space).
We discuss the kinematic properties of these structures, and cross match them with APOGEE~DR17 and LAMOST~DR6 to study whether many of the small clusters belong to a similar larger cluster based on their chemical abundances. 
Although we do not identify any new structures, we find that the \HDBSCAN~member selection of already known structures is unstable to input kinematics of the stars when resampled within their uncertainties.
We therefore present the most stable subset of local kinematic structures, which are consistently identified by the clustering algorithm, and emphasize the need to take into account error propagation during both the manual and automated identification of stellar structures, both for existing ones as well as future discoveries.

\end{abstract}


\section{Introduction}

In the most popular cosmological structure formation model, typical spiral galaxies, like our Milky Way, form hierarchically through many mergers with less massive galaxies \citep[see e.g.][]{White&Rees_1978,frenk12,wechsler18,behroozi19}. Theories and simulations predict that most of the merging stellar populations are tidally disrupted during infall, and form various structures in today's Milky Way stellar halo \citep{zolotov09,cooper10}. These accreted stars appear in large all-sky surveys as spatially coherent kinematic structures (``streams''), as well as mere overdensities in orbital energy and angular momenta after phase mixing (``tidal debris'') (for a review see e.g., \citet{helmi20}). Studying these accreted stars allows us to estimate properties of their progenitors (e.g., mass, age, infall time, and kinematics). The stellar chemical signatures of such structures, such as metallicity and $\alpha$-element abundances \citep{freeman02,frebel15,maiolino19,sestito19,fattahi20}, and kinematic properties \citep{johnson96,johnson98,helmi99b,bullock05,johnston08} thus provide unique information about the formation history of the Milky Way. 

Understanding the merger history of the Milky Way and the properties of the accreted and merged progenitors galaxies can help us learn about near-field cosmology and provide constraints on structure formation and galaxy evolution in the late time universe (see e.g., \cite{myeong19,2019MNRAS.482.3426M,belokurov20,2022arXiv220603875F}). At the same time, \citet{2018PhRvL.120d1102H,2018JCAP...04..052H,necib19a} found that the stellar component of the accreted systems may be used as a tracer of the accompanying dark matter (DM) accreted along with the system.  
The DM distribution in the Milky Way, especially near the solar neighborhood, is one of the key ingredients towards uncovering the particle nature of DM. Direct detection rates of DM particles rely on the proper modeling of the local DM density and velocity distributions, which translates discoveries as well as null results into a DM mass and DM-nucleon scattering cross section \citep{Goodman85,Drukier86,1996PhR...267..195J}.

Using accreted stars to answer near-field cosmology questions and as DM tracers relies on efficiently and accurately identifying them in the local solar neighborhood. Since our solar system resides in the thin disk of the Milky Way, the local stellar population is dominated by in-situ stars formed in the thin disk \citep{soubiran03}. Thus, while earlier discoveries and studies (see e.g., \cite{ibata94,belokurov06,schlaufman09}) were carried out with chemical abundance measurements and small kinematic data sets, identifying some of the more elusive accreted stellar populations, such as Thamnos and the metal-weak Atari disk, on top of this in-situ background relies on obtaining 6D kinematic information for a large data set of local stars. 
Large kinematics data sets are powerful in identifying intact satellites that are in the early stage of accretion, diffuse satellites that have phase-mixed into the environment, as well as those that are in the process of transitioning from the former to the latter via tidal disruption. This can now be done with large photometric and astrometric surveys such as the \Gaia mission \citep{gaia16,gaia18,gaia21}. 

Many studies have already used these data to identify numerous substructures that merged into the Milky Way in the past \citep{belokurov18,helmi18,myeong18a,koppelman19a,necib19a,helmi20}. Among these local structures, the Gaia-Enceladus/Sausage (GSE) is the most recent major merger event experienced by the Milky Way \citep{belokurov18,helmi18}. Additionally, \citet{koppelman19a} discovered the Thamnos structure in the \Gaia second data release (DR2) as part of the local retrograde halo. Sequoia, a high energy retrograde structure originally attributed to being part of the GSE by \citet{helmi18}, was found to be most likely coming from a separate accretion event in \cite{myeong19} with \Gaia DR2 data. Sharing similar kinematic properties, Arjuna and I'itoi were discovered in \citet{naidu20} as distinct accretion debris given their metallicity differences after combining data from the \Gaia DR2 and the H3 survey\citep{gaia18,conroy19}. Wukong and Aleph, two prograde structures, were found similarly by \citet{naidu20}: Aleph is a highly prograde structure with circular orbits and a significant vertical motion, as well as a metal-rich and relatively alpha-poor chemistry. Wukong is also a prograde structure lining the margin of the GSE in the $E_{tot}~vs.~L_z$ plane, shown as two over-densities at $E_z \sim -1.1$ and $-1.3 \times 10^5$\,km$^2$\,s$^{-2}$. Spatially coherent streams such as S1, Icarus, and Nyx, to name a few, are also identified \citep{myeong18a, meingast19, refiorentin20}. They are associated to either past merger events or disk clusters. 

Past discoveries of local over-densities such as the Helmi stream \citep{helmi99a} have also been further studied with \Gaia data \citep{koppelman19b}. The long known ``metal weak thick disk" \citep{Norris85,Morrison90,chiba00} has recently been postulated to be likely the result of a merger, rather than part of the canonical thick disk. \citet{mardini22} confirmed such origin with photometric metallicities from the SkyMapper survey \citep{keller07,wolf18,chiti21} and kinematics from \Gaia. Using machine learning techniques trained on simulations \citep{2016ApJ...827L..23W,2018MNRAS.480..800H,2020ApJS..246....6S}, and labeling the \Gaia DR2 data set, \cite{2020A&A...636A..75O} built an accreted star catalog \citep{ostdiek_bryan_2019_3579379} from which multiple structures were identified \citep{2020ApJ...903...25N}, including Nyx \citep{2020NatAs...4.1078N}, a stream within the thick disk with a highly eccentric orbit. 

These studies use the fact that accreted substructures emerge as overdensities in the 6D phase space. The key, thus, is extracting such overdensities effectively and robustly from the data. 
Current studies mostly rely on two main methods in identifying overdensities and selecting their constituents: manual selection \citep{helmi18,2019ApJ...885..102L,naidu20,2021ApJ...908..191C} and clustering algorithms \citep{helmi17,necib19a,necib19b,2019ApJ...880...65Y,wu21,Lovdal22}, or some combination of both \citep{koppelman19a,2021ApJ...919...66B} (see \citet{buder22} Table~A.1 for a review of techniques). 

For manual selection, overdensities in phase space are identified based on physical intuition gained from past discoveries \citep{naidu20} or statistical comparisons with some reference random data set \citep{helmi18}. They are then characterized based on their phase space properties. 
Potential background contamination is removed by examining stellar population properties, such as age and metallicity distributions. 
This method relies on the fact that an overdensity is located in phase space where the background contamination level is low, either because the substructures themselves have peculiar properties which set them apart from the majority of the in-situ Milky Way stars, or because astronomers are able to apply well-informed cuts on the data set to remove most of the background. For structures that reside within or close to the Milky Way background, manual selection becomes difficult and tricky. 

For machine learning-focused methods, structures are picked out directly from the data sample with minimum preprocessing by various algorithms. The results are, in general, more consistent between works, for example, especially when using the same algorithm. Naturally, this is not frequently true for manual selections. In most cases, however, the final results need to be further examined, 
combined with additional less complete information such as metallicity and chemical abundances, to remove unphysical structures such as background noise and data artifacts. \cite{kaley22} tested kinematic-based clustering algorithms on simulations and showed that they generally lack efficiency and accuracy in identifying structures resulting from now-dissolved ultra faint dwarf galaxies. 

Despite the difficulties mentioned above, both methods have yielded many new structures in the solar neighborhood that have been extensively studied and validated. 
Yet, it is still often the case that the stellar membership of any of these structures has to be confirmed with archive/follow-up spectroscopic observations if they were initially discovered based on kinematic information only. Even with the best informed physical expectation and/or the most sophisticated algorithms, the kinematic selection alone is typically not enough to robustly determine the constituent stars, even when the overdensities/structures are principally well-established. This is due not only to the background-dominated signal search (i.e., the in-situ population) but also, and more importantly, due to our inability to measure the stellar kinematics infinitely precisely. Despite the recent advancement in astrometric survey precision, uncertainties still play a large role that affects stellar membership assignments. For practical reasons, uncertainties are often left out in kinematic structure studies, either because the machine learning algorithm inherently does not support uncertainties or because the process becomes too computationally intensive. This lack continues to hamper the robustness of identifying additional structures to eventually identify all progenitors of the Milky Way.

In this paper, we aim to incorporate uncertainties for the first time into an unsupervised machine learning algorithm to search for Milky Way substructures in \emph{Gaia}. The goal is to select a robust sample of star members present in clusters/overdensities purely based on kinematic inputs. We use the Hierarchical Density-Based Spatial Clustering of Applications with Noise (\HDBSCAN) \citep{mcinnes2017hdbscan}, an unsupervised learning algorithm, to identify clusters in the 6D phase space with data from the \Gaia eDR3 data set \citep{gaia21}. We combine Monte Carlo resampling and Jaccard coefficients to estimate the robustness of the \HDBSCAN~clusters. Doing so currently leads us to identify only the most robust clusters and their member stars. We recover the GSE, Sequoia, the Helmi Stream, and globular clusters NGC~3201 and NGC~104, but do not Thamnos and Nyx due to an initial cut on the maximum orbital vertical distance \zmax, or Arjuna and I'itoi which could be a subset of Sequoia in our analysis, or Aleph and Wukong that we expect to be either unstable, or dominant further away from the disk than our sample. Our results demonstrate the importance of incorporating uncertainties in machine learning, especially unsupervised clustering, for astronomy studies, and how this approach affects our interpretation of the clustering results and the constituent stars within the clusters. 

This paper is structured as follows: We discuss the data sample used for this study in Section~\ref{sec:dataset}. The clustering algorithm is discussed in Section~\ref{sec:clustering_algorithm}. We present the clustering results in Section~\ref{sec:cluster_res} and discuss their interpretations in Section~\ref{sec:discussion}. 

\section{Data Set}
\label{sec:dataset}

\subsection{Quality Cuts}

We first apply quality cuts on \Gaia eDR3 stars, combined with the radial velocity measurements from \Gaia DR2. 
Based on recommendations from \cite{lindegren21a}, we select sources with good astrometric solutions by requiring \texttt{parallax\_over\_error > 5}, \texttt{ruwe < 1.4}, \texttt{astrometric\_excess\_noise < 2}, \texttt{duplicated\_source = 0}, $G < 19$, and \texttt{visibility\_periods\_used >= 10}. We also perform the cut on color \texttt{C = phot\_bp\_rp\_excess\_factor} with the prescription described by \citet{riello21} to remove faint sources in close proximity to bright sources. 
We then correct for the zero point bias in the parallaxes as described in \citet{lindegren21}.

The data sample size, after applying the quality cuts mentioned above, and removing high-velocity stars ($|v_{i}| > 1000$~km\,${\text{s}^{-1}}$ where $i = r, z, \phi$), is $5,557,670$. More cuts are applied to the sample before going through the clustering algorithm, which are described in the following sections.

\subsection{Crossmatched Data Sets}
\label{crossmatch_datasets}

For extra chemical information, we crossmatch the sample with the spectroscopic surveys LAMOST DR6 (632,223 matches) \citep{cui12,zhao12} and APOGEE DR17 (211,798 matches) \citep{majewski17,abdurrouf22}. 
We focus mainly on the metallicity measurements from these two surveys, as LAMOST provides the most matching sources and APOGEE provides the most $\alpha$-abundances. 

Chemical abundance information is used for characterization of individual clusters, comparison between clusters, and comparison with literature. To avoid systematic differences across data sets, we limit metallicity characterizations of our identified clusters to one survey at any given time. Unless otherwise noted, we use APOGEE metallicities and $\alpha$ abundances in summary plots for characterizing individual clusters as APOGEE chemical abundances are derived from high-resolution spectra than those from LAMOST (medium-resolution spectra only). However, for examining relationships between the final clusters in this study and for a comparison with known structures' metallicities from the literature, 
we instead use metallicities from LAMOST that are available for more of our cluster stars and thus provide better statistics. We discuss more details in Section~\ref{sec:discussion}. 

\subsection{\zmax~Cuts}

\begin{figure*}[t]
\begin{center}
\includegraphics[angle=0,width=0.9\textwidth]{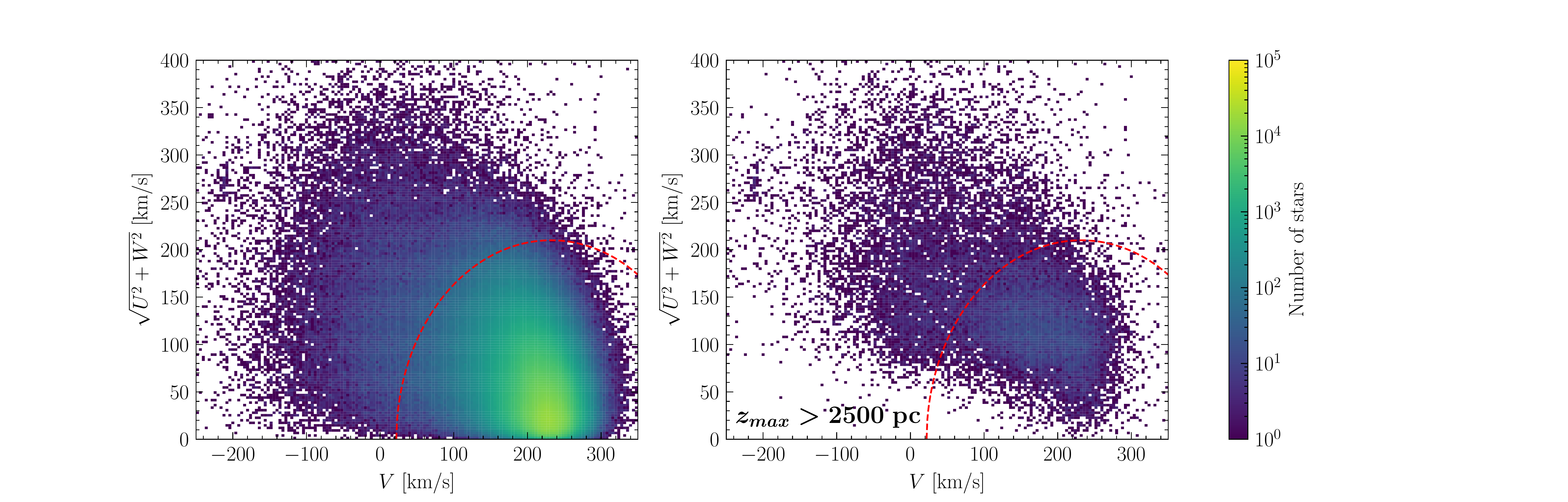}
\end{center}
\caption{Toomre plots for the full sample (left) and the subsample with \zmax~cut at 2500 pc, as indicated at the bottom-left of each panel. The red dashed line indicates a probable LSR velocity cut at 210 km\,${\text{s}^{-1}}$, with $V_{LSR} = 232$\,km\,${\text{s}^{-1}}$.}
\label{toomre_plots}
\end{figure*}

\begin{figure}[t]
\begin{center}
\includegraphics[angle=0,width=0.45\textwidth]{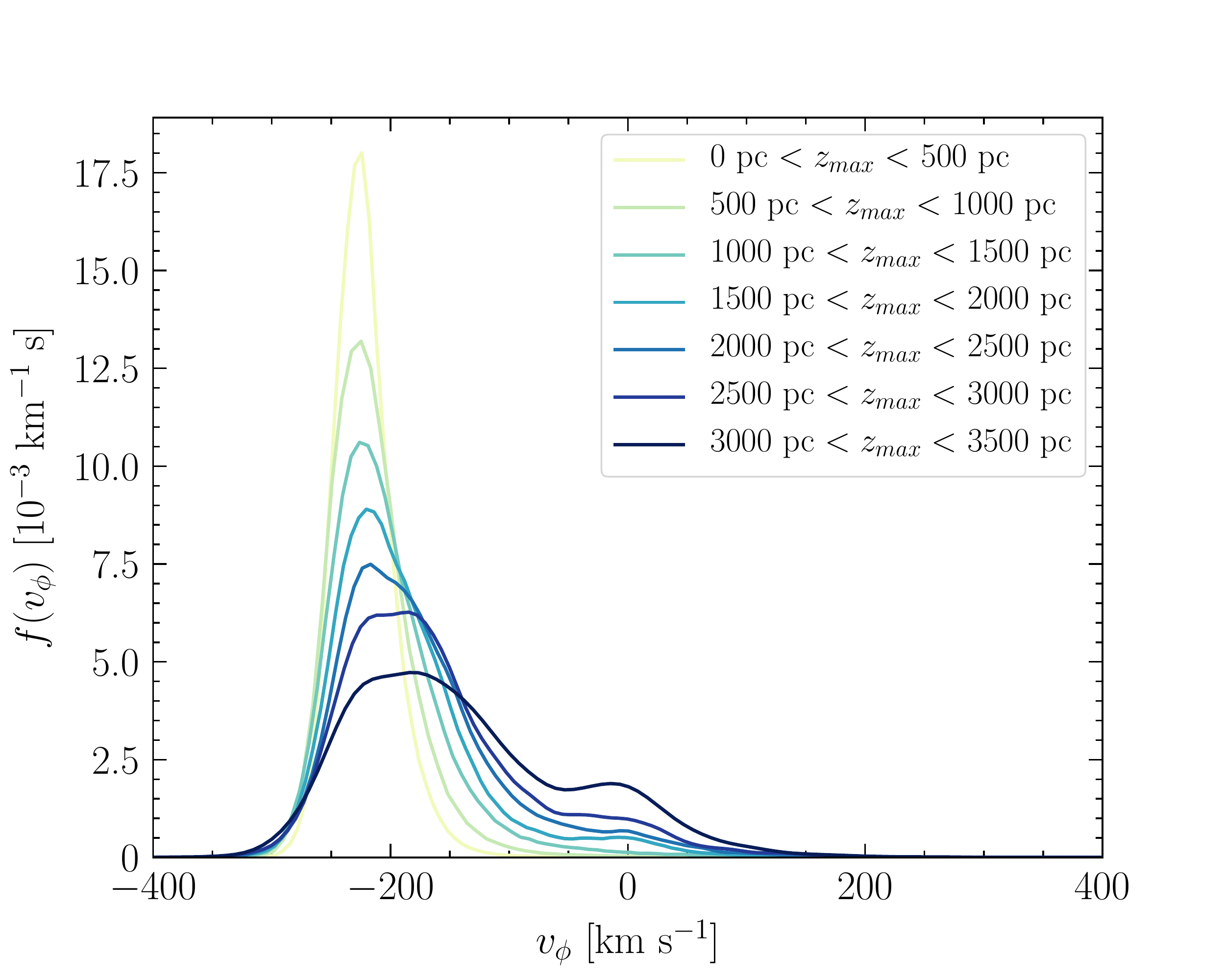}
\end{center}
\caption{Normalized distribution of \vphi~for increasing \zmax~intervals, including only stars with available \Gaia2 RV. The \vphi~histogram was constructed from bins between 0 to 3.5 kpc, in 500 pc intervals. There is a gradual shift in the overall distribution away from the LRS at $\sim-$220~km\,${\text{s}^{-1}}$ towards 0 km\,${\text{s}^{-1}}$, as the sample evolves from thin-disk dominated to halo dominated.
\label{zmax_int_summary}
}
\end{figure}

It is extremely difficult, if not impossible, for most clustering algorithms to identify substructure signals over a strong background. The majority of our sample is made up of in-situ thin disk stars, which, if not removed in some way, will flood all signals and prevent detecting any meaningful clustering. To remove the thin disk contamination, a standard practice in the field has been to perform a velocity cut in the local standard of rest (LSR) velocity (see e.g., \cite{helmi18,Lovdal22}). Where spectral information is available, velocity cuts are also combined with metallicity cuts to exclude metal-rich disk stars (see e.g., \citet{myeong18a}). This method is the most successful in completely removing stars with thin disk-like kinematics, thus leaving behind a sample comprised primarily of halo stars and some hot thick disk stars. However, the disadvantages of this method are that it might bias the overall velocity distribution of the resulting sample, or drop some of the prograde structures (e.g., Nyx \citep{2020NatAs...4.1078N}). Since our goal is to identify velocity structures and eventually their impact on the local DM velocity distributions in future work, we here opt for the least kinematically biased sample possible by instead placing a \zmax~cut rather than using a direct velocity cut. 

Specifically, we apply a cut at \zmax$> 2.5$ kpc to remove the thin disk, and to keep the possibility of studying potential prograde structures with high vertical motion. 
The left panel of Fig.~\ref{toomre_plots} shows the Toomre plot of our initial sample, where stars in the thin disk ($V\sim220$ km\,${\text{s}^{-1}}$) dominate by orders of magnitude over the rest of the stars, which would also include any prograde kinematic substructures.

Our \zmax~is calculated through orbit integration, as detailed in Section \ref{sec:orbits}.
We expect most in-situ stars to have \zmax~no higher than 2.5 kpc, especially with a thin disk scale height approximately at 300 pc \citep{juric08}. Thus, this cut will serve the purpose of removing the thin disk, similar to the LSR velocity cut, but at the same time it will allow potential prograde structures with high vertical motion to remain in the final sample. As shown in Figure~\ref{zmax_int_summary}, the \vphi~distribution of stars at higher \zmax~interval shows a clear transition from the expected rotational velocity of the disk of $-$220~km\,${\text{s}^{-1}}$ towards 0 km\,${\text{s}^{-1}}$, the average rotational velocity of the stellar halo. Figure~\ref{toomre_plots} also indicates a significant portion of the stars that would have been removed by an LSR velocity cut (red dashed line) remaining in the post-cut sample. Our \zmax~cut is not as clean as a LSR velocity cut, by design, as even at vertical distances a few times above the scale height of the thin disk, the exponential tail of the thin disk is still comparable in size to the slower rotating prograde halo, and retrograde halo populations. \citet{juric08} measured the local halo-to-thin disk stellar number density ratio of 0.5\%. 
In addition to that, the uncertainty in our distance measurements unavoidably creates a systematically puffed-up thin disk, causing a significant amount of prograde stars with kinematics very similar to that of the thin disk even at \zmax~greater than 3~kpc. Yet, doing so gives us the opportunity to study potential overdensities in the high vertical motion tail of the thin disk. 

More specifically, in terms of potential velocity biases, while this cut still poses an implicit bias on the distribution of the velocity in the $z$ direction, it is nevertheless much less restrictive than the LSR cut. Again, this leaves space for prograde accreted stars to be part of the clustering sample. Following this reasoning, we require \zmax$> 2.5$~kpc, as well as that the mean value of \zmax~has to be larger than two standard deviations above the cut when taking into account the errors in the measurements. 

\subsection{Coordinate Transformations}

After sample selection, we compute the galactocentric coordinates and velocities assuming the \Astropy~galactocentric frame with default parameters adopted in \Astropy~v4.0 \citep{2013A&A...558A..33A,2018AJ....156..123A}. The galactic center coordinate is at ra=266.4051\degree~and dec=-28.936175\degree \citep{reid04}. The right-handed coordinate system is defined such that the x-axis points from the Sun to the galactic center and the y-axis points in the direction of local disk rotation.
The distance to the galactic center is assumed to be 8.122~kpc \citep{gravity18}, with the Sun moving with velocity (12.9, 245.6, 7.78)~km\,${\text{s}^{-1}}$ and located 20.8~pc above the midplane \citep{drimmel18,bennett19}. 

Uncertainties on the galactocentric velocities and positions are propagated from uncertainties in parallax, proper motion, and radial velocities. We take into account covariances between \Gaia measured parallax and proper motions. As a result, there exist non-trivial covariances between the transformed galactocentric velocities, which we will take into account in the orbit integration. Radial velocities are measured from \Gaia DR2 and have no provided covariances with other \Gaia EDR3 astrometry measurements.

\subsection{Orbit Integration}
\label{sec:orbits}

We integrate the orbits for our final sample backward in time for 2.5~Gyr with 1~Myr steps using the Python \texttt{agama}~package~\citep{vasiliev19}. We set up the potential to be the default \texttt{MilkyWayPotential}~from \textsc{gala}~v1.4.1, which is a four-component potential model consisting of a bulge and nucleus with Hernquist profile \citet{hernquist90}, a disk with Miyamoto-Nagai profile \citet{miyamoto75}, and a halo following the NFW profile \citet{navarro97}. The disk model is taken from \citet{bovy15} and the rest are fixed by fitting to a compilation of mass measurements of the Milky Way. More details on the parameters can be found in the \textsc{gala}~documentation \citep{gala}. Orbital parameters including energy, angular momenta, eccentricity, maximum galactic $Z_{\rm{max}}$, apogalacticon, perigalacticon, and actions are derived from the integrated orbits.

For each star, we estimate the uncertainties for the orbital parameters by resampling the initial positions and velocities 100 times with their respective uncertainties (assuming the uncertainties are inherently Gaussian) and including the covariances between the velocities. Actions are calculated using the \texttt{ActionFinder} routine provided by \texttt{agama}. 
We note that while their effect is minimal, we include the covariance between the positions and velocities to compute the uncertainties in the orbital parameters.
We then record the mean and standard deviation of these parameters as the final values and corresponding uncertainties.

\section{Clustering Algorithm}
\label{sec:clustering_algorithm}

The main goal of this study is to identify kinematic clusters, from machine learning algorithms, which are stable under resampling within the uncertainties of the stellar kinematics inputs.
Our method thus focuses on ensuring that the members of the extracted clusters are stable against being assigned as noise or as parts of different clusters as can occur when uncertainties are not taken into account. Consequently, stars we assign as cluster members are very stable against any splitting/merging of their apparent host clusters; splitting refers to a cluster breaking into smaller clusters, and merging refers to multiple clusters merging into one single cluster.

Given the complexity of this section, we summarize the main points here. We apply the clustering algorithm \HDBSCAN~(defined with its parameters in Section \ref{sec:hdbscan}) on two separate spaces: The 3-dimensional cylindrical velocity space, and the 3-dimensional action space with total energy as an additional axis (see Section \ref{sec:clustering_spaces}). In order to incorporate the uncertainties, we first define a high-stability sample by only selecting the stars that get clustered more than \Nstackcut~times (defined in Section \ref{sec:nstackcut}) when applying \HDBSCAN~over 100 realizations, as shown in Section \ref{sec:uncertainties}. We then cluster the high-stability stars, and define the Jaccard coefficient as a measure of the stability of clusters (Section \ref{sec:stability}).

\subsection{Choice of Algorithm}
\label{sec:hdbscan}

We adopt \HDBSCAN\footnote{\url{https://hdbscan.readthedocs.io/}} as a clustering algorithm for identifying the kinematic substructures in the \Gaia data \citep{mcinnes2017hdbscan}. The choice of the clustering algorithm depends on the particularity of the data set. Specifically, we do not know a prior total number of clusters, the clusters could be of various shapes, and the data inherently has stars that are not part of a cluster, and would therefore need to be treated as noise. Ideally, the required algorithm would address the specifics of the data above, and potentially it would incorporate measurement uncertainties. Given the difficulty in finding an algorithm that has a standard treatment of measurement errors, a stability study of the resulting clusters is required, as we will discuss in Section~\ref{sec:stability}. 

\HDBSCAN~stands out to be the most ideal choice for our study. The algorithm transforms the space according to the density of points, and performs single linkage clustering on the transformed space. \HDBSCAN~assumes dense regions as clusters and thus makes no assumption about the intrinsic distribution or total number of clusters. It also does not require every data point to be assigned to a cluster, allowing the presence of noise. It builds a merger tree, where one can cut at variable heights to decide the smallest sized cluster based on a few hyperparameters, mostly \texttt{min\_cluster\_size}, which determines whether a split of the tree should be treated as two new clusters or one cluster losing noise points. The resulting clusters can thus have varying densities as long as they are stable against further splitting. Additionally, the clusters found are deterministic with the same input data and hyperparameters setting. 
\citet{kaley22} examined the performance of a suite of clustering algorithms on the Caterpillar cosmological simulation \citep{griffen16} while trying to recover substructures left behind by accreted ultra-faint dwarf galaxies. Their result demonstrates that \HDBSCAN~has the best performance among all algorithms, albeit not powerful enough to identify all subtle, long-accreted clusters. 

The parameters of the algorithm are set to default with the exception of \texttt{min\_cluster\_size}, which is set to 20 to capture the smallest substructure, and \texttt{cluster\_selection\_method}, which is set to \texttt{leaf} to track the leaf nodes of the clustering tree. As described in Section~\ref{sec:stability}, the merging of these smaller homogeneous clusters along the hierarchical tree can be altered significantly due to uncertainties. Instead of having \HDBSCAN~decide how the leaf nodes should be merged, it is more consistent to first examine the stability of the leaf nodes individually (and how they change with uncertainties folded in),
as shown in Section~\ref{sec:stability}. We then considered separately if and how they may be merged, as shown in Section~\ref{sec:cluster_relation}.

\subsection{Clustering Spaces}
\label{sec:clustering_spaces}

In this work, we apply \HDBSCAN~to two separate spaces: The 3-dimensional cylindrical velocity space and the 3-dimensional action space with total energy as an additional axis. First, the cylindrical velocity clustering space is chosen as it will identify stellar kinematic substructures, which might correlate with DM velocity substructures.
As shown in \cite{necib19b}, the stellar velocity distribution of debris flows would be correlated with the velocity distribution of DM, which is key in studying DM direct detection rates. 

Second, we also perform clustering in the integrals of motion space for accreted structures that have phase-mixed since their accretion. In particular, we perform clustering on the total energy and the 3D action space. Previous works have discussed the fact that accreted stellar populations remain clustered in such integral of motion space long after the merger, given a slowly varying potential, from both theory \citep{binney82} and simulation \citep{helmi99b,kenbe05,gomez10} perspectives. While these structures may no longer show a coherent velocity distribution, they are nonetheless important sources of accreted DM and would affect the local DM velocity distribution. 

For the reasons mentioned above, we do not expect a one-to-one correspondence between the clusters identified in one space and those from the other. Nonetheless, we compare the clustering results from these two spaces in Section~\ref{sec:cluster_res} as these results complement each other. Moreover, they allow us to study, in future work, the local DM velocity distribution from both coherent and phase-mixed accreted structures.

\subsection{Uncertainties: Random Realizations and Stacking}
\label{sec:uncertainties}

\HDBSCAN, similarly to most clustering algorithms, is not designed for handling data with uncertainties. Previous work (see e.g., \citealt{koppelman19a}) that had used \HDBSCAN~for cluster identification, did not take uncertainties into account when applying the algorithm to kinematic data. Other works have included measurement uncertainties on a much smaller scale, typically on the order of thousands of stars; for example, \citet{limberg21,gutin21,shank22} used \HDBSCAN~to identify dynamically tagged groups for very metal-poor stars. \citet{limberg21} proposed estimating the confidence level of the clusters via feeding Monte Carlo generated perturbed data sets into the \HDBSCAN~hierarchical tree generated from the nominal values. The cluster membership assignment is re-evaluated using the prediction method provided by the \HDBSCAN~package.\footnote{This procedure provides estimates on the probability of a given star being the member of a cluster, considering uncertainties on the kinematic properties, which indirectly evaluates the robustness of the clusters themselves. All probabilities are based on the hierarchical tree generated by \HDBSCAN~from the nominal (or mean) kinematic properties of the sample stars.} As pointed out by the authors of \HDBSCAN, the algorithm is a transductive method, meaning new/perturbed data points can and should alter the underlying clustering \citep{mcinnes2017hdbscan}. This is especially the case if one attempts to predict the membership assignment of an entire perturbed data set with a size identical to the original nominal data set used to generate the tree. The membership assignments for the perturbed data sets are thus inherently inconsistent with the hierarchical tree. 

In this work, we adopt a self-consistent method of simultaneously evaluating the robustness of the clusters themselves and the membership of the stars within the clusters by regenerating the hierarchical clustering tree for each Monte Carlo resampled data set.  
Specifically, uncertainties are incorporated by repeating the clustering process with input parameters resampled from their respective uncertainties for 100 times. 
We repeat the clustering procedure with 100 realizations, sampled from the error bars on each of the variables included. The clustering results from each random realization are then combined together in order to establish the stability of the clusters. We find that, while the general clustering results may appear similar by eye in various spaces, the exact stellar membership could differ significantly between realizations. Clusters that are in close proximity to other clusters often exchange members due to the existence of noise and background. Such behavior results in the clusters themselves shifting around in the clustering space, and sometimes merging and splitting between different realizations. It is thus extremely difficult to directly quantify the stability of the clusters.

To overcome this challenge, instead of directly evaluating the stability of individual clusters, we first focus our attention on the stars themselves. We stack the clustering results from all 100 realizations, and instead of recording the membership of the stars to specific clusters (or noise), we simply record whether or not the stars are identified to be part of some cluster at all. The result is a distribution of the number of times a given star is clustered to be non-noise. Stars that are consistently clustered can then be selected by making a cut on the number of times the star has been clustered; we call this variable \Nstack. An \Nstackcut~allows us to identify the stable stars first, that are more often part of a cluster and not the noise, which can then be used to find the stable clusters. We call the sample of stars after this cut the high-stability sample. 

\begin{figure*}
\includegraphics[angle=0,width=0.95\textwidth]{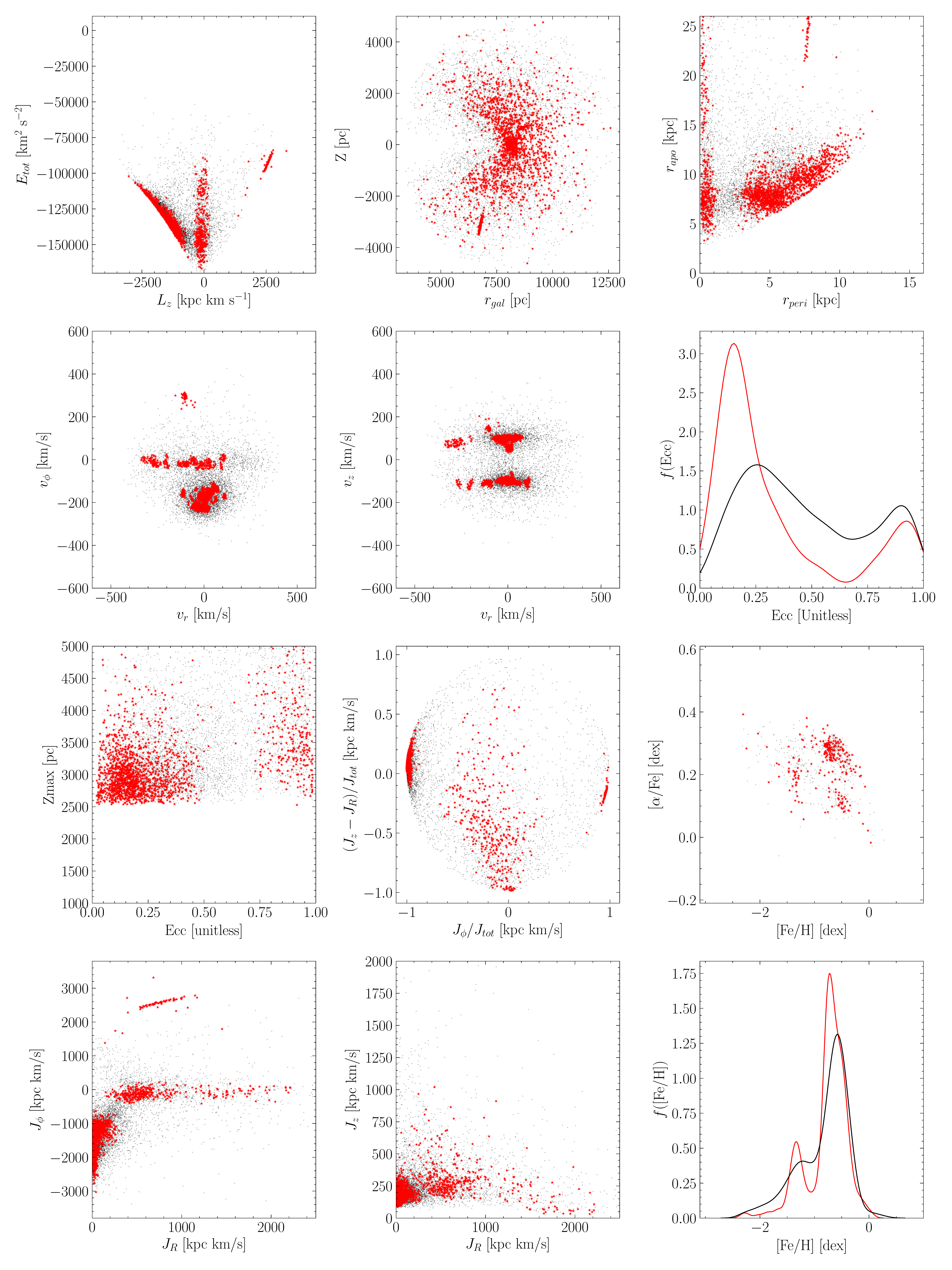}
\caption{Summary plot for the full sample (black), downsampled to 20\%, with the high-stability (\Nstack$>20$) sample (red) identified from the cylindrical velocity clustering space.}
\label{vel_stack_20_sample}
\end{figure*}

The sample size of the high-stability sample is significantly smaller than the full sample. It does, however, guarantee that the clusters found in this sample are much more robust with respect to any resampling of the kinematic inputs within their uncertainties. 
The fact that these stars are associated with some cluster in multiple random realizations signifies that these stars stay close to overdensities in the parameter space. More importantly, the overdensities themselves remain significant above the background such that \HDBSCAN~would identify them as clusters. We can think of this as deepening the contrast of the clustering space. We show these high-stability sample stars (in red) as well as the entire sample (in black) in Figure~\ref{vel_stack_20_sample}. Some structures are visible by eye in this high-stability sample.

\subsection{Selection of \Nstackcut}
\label{sec:nstackcut}

We choose \Nstackcut=20 to balance sample sizes and overall stabilities of the final clusters. The sample size of the high-stability sample reduces as shown in Figure~\ref{vel_stack_Nstack_dist}, while the stabilities of the final clusters improve with increasing \Nstackcut. We tested the clustering results' stabilities of high-stability samples using different \Nstackcut, following methods detailed in Section~\ref{sec:stability}, and found that \Nstackcut~at 20 yields the largest sample of stable stars with most of the final clusters being stable.

We find no obvious correlation between the \Nstackcut~and the uncertainty distribution of the high-stability sample. A concern of the \Nstackcut~would be that it reduced the data size by essentially performing an uncertainty cut, biasing towards stars that do not move as much in the clustering space. To test this, we examine the uncertainty distribution of the subsample in 3D cylindrical velocity space with and without \Nstack~cut, shown in Figure~\ref{vel_stack_err_dist}. The result indicates a minimum effect on the overall distribution of the uncertainties in any of the cylindrical velocities. We are thus confident that the stars with lower \Nstack~are not excluded because of high uncertainties but because they are at the edge of the identified clusters in the clustering spaces. In other words, under the same level of uncertainties, they blend in with the background noise more easily given their position in the clustering space, and thus \Nstack~is selecting physically more significant stars as part of a cluster than just statistically better measured stellar kinematics. The same test is carried out in the integrals of motion space with a similar result.

After identifying the stable stars, we apply \HDBSCAN~with the same parameter settings on only the stars, with \Nstack~above 20. The resulting clusters on this high-stability sample are taken as the final clusters.

\makeatletter\onecolumngrid@push\makeatother
\clearpage

\clearpage
\makeatletter\onecolumngrid@pop\makeatother

\subsection{Stability of Clusters and Stars}
\label{sec:stability}

We apply the clustering algorithm as well as the 100 random realization process mentioned above to this new high-stability sample. As discussed in Sections~\ref{sec:uncertainties} and \ref{sec:nstackcut}, the high-stability sample comprises only stars identified to be part of some clusters in at least 20 of the 100 random realizations. The clusters found this way are thus more robust than those from the full sample. They are also more stable against merging/splitting due to uncertainties because stars consistently identified as noise have been removed. We now discuss in more detail how we evaluate the stability of the final clusters.

We call the clusters found from the mean values of the clustering parameters the baseline clusters. 
We use the Jaccard coefficient $J$ \citep{jaccard01} between two clusters from two different realizations to quantify the stability of the cluster. The coefficient $J_{s}^{a}$ for a baseline cluster $s$ is defined as the number of common members between the cluster $s$ and a cluster $a$ found in another realization, which we label $N_{s,a}$, divided by the total number of members in the initial cluster $s$ that we label $N_s$.

\begin{equation} \label{eq:jaccard}
    J_{s}^{a} = \frac{N_{s,a}}{N_{s}}.
\end{equation}

Baseline clusters $s$ are those from the baseline realization, i.e., clusters found with the mean values, and clusters $a$ are those found in each of the 100 random realizations.\footnote{For illustrative purposes, we compute $J_{s}^{a}$ between all baseline clusters and random realization clusters, i.e., for every baseline cluster, we study how much it overlaps with every other cluster in all 100 random realizations. Theoretically, the same thing could be done with any one of the random realizations serving the role of the baseline clusters, because they are ultimately treated as equal. The conclusion about the stability of the cluster membership remains valid, however, regardless of whether we use the baseline clustering as the base, or a random realization as the base. We use the mean values clustering result as the baseline. The only case of clusters that we might miss are the ones that are present in other realizations but not in the baseline.}

We also define the inverse Jaccard Coefficient $J^\dagger$ as the number of common members divided by the total number of members in a random cluster. More specifically, the $J_{s}^{a\dagger}$ for a baseline cluster $s$ and a random realization cluster $a$ is therefore defined as

\begin{equation} \label{eq:inverse_jaccard}
    J_{s}^{a\dagger}= \frac{N_{s,a}}{N_{a}}.
\end{equation}

To study whether a baseline cluster $s$ is stable, we find its closest cluster in a random realization, i.e., the cluster with which it shares the most members. To do so, we calculate the maximum $J_{s}$, that we call $J_{s}^{r, \rm{max}}$, of the baseline cluster $s$ (with respect to all the clusters) in each of the random realization $r$. We thus record at most 100 values of the $J_{s}^{r, \rm{max}}$ for each baseline cluster $s$; some baseline clusters are not identified at all in some of the random realizations, in which cases the $J_{s}^{r, \rm{max}}$ is undefined. We also record the corresponding maximum value of the $J_{s}^{\dagger}$ in each realization, called $J_{s}^{r\dagger, \rm{max}}$; this is applying Equation \ref{eq:inverse_jaccard} on the same pair of clusters (the baseline $s$ and the random realization $r$). We note that these $J_{s}^{r\dagger, \rm{max}}$ are therefore not necessarily the maximum possible values for the given random realization and baseline cluster.\footnote{For example, in the case of baseline clusters splitting into multiple smaller clusters in random realizations ($J_{s}^{r\dagger, \rm{max}} \neq J_{r}^{s, \rm{max}}$).} We examine the $J_{s}^{r, \rm{max}}$ and $J_{s}^{r\dagger, \rm{max}}$ distribution of all the baseline clusters to evaluate their stability.

\begin{figure}
    \begin{center}
    \includegraphics[angle=0,width=0.45\textwidth]{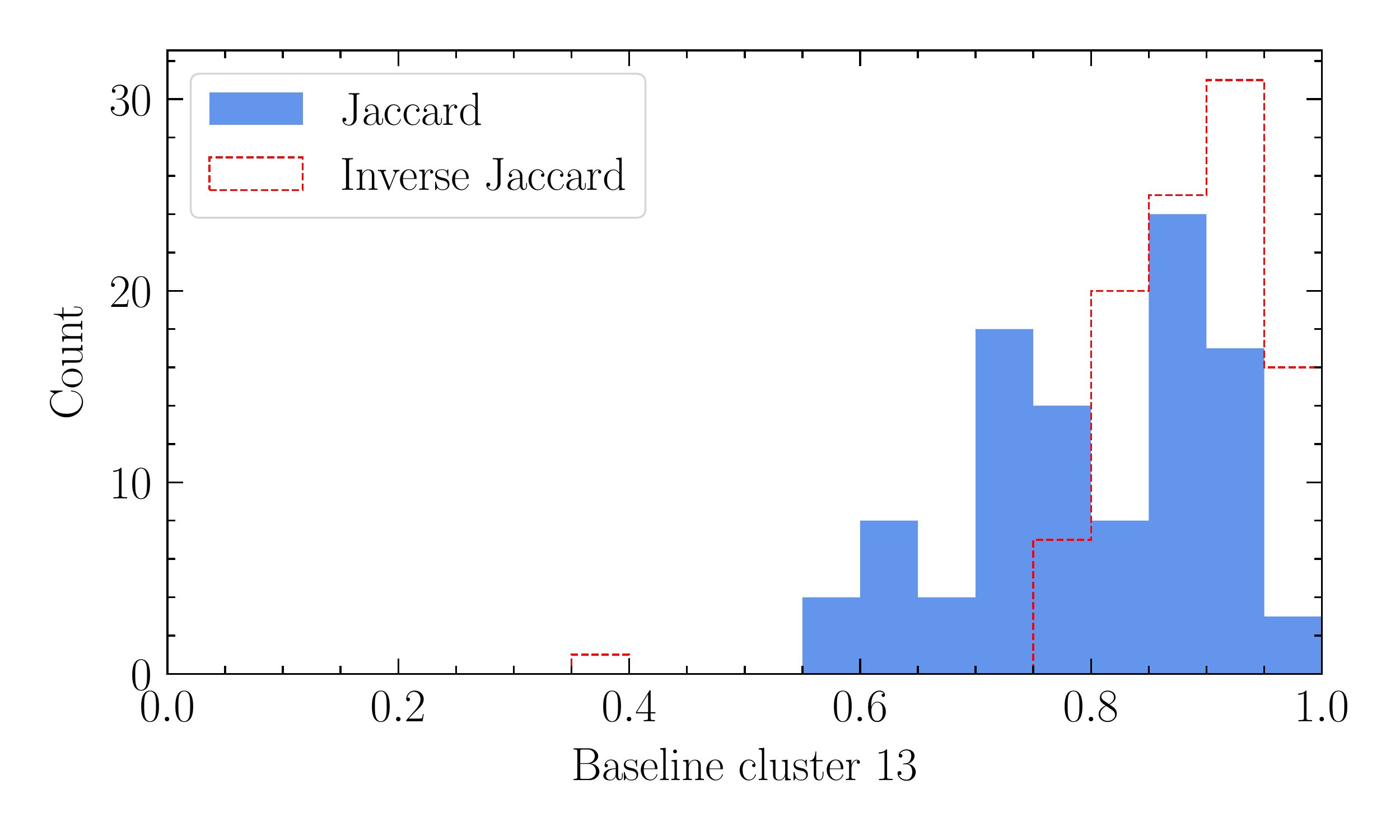} \\
    \includegraphics[angle=0,width=0.45\textwidth]{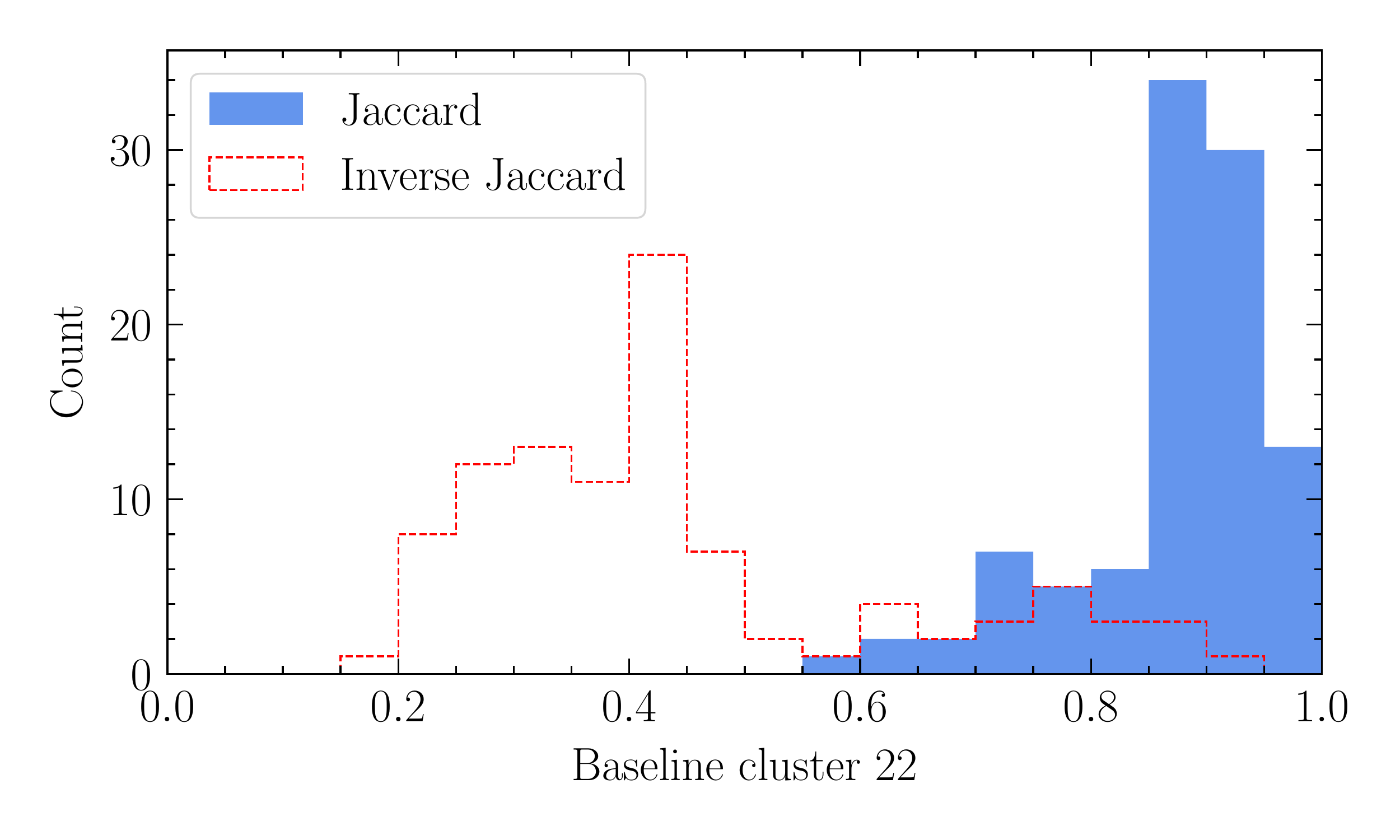}
    \end{center}
    \caption{Example Jaccard and inverse Jaccard distributions from clustering results in cylindrical velocity space with \zmax~cut at 2.5~kpc and \Nstack~cut at 20. Cluster 13 represents a typical stable cluster, whereas cluster 22 represents a more ambiguous case where the Jaccard distribution seems stable while the inverse Jaccard is not. See text for detailed discussion of this case.}
    \label{fig:jc_example}
\end{figure}

To distinguish stable from unstable clusters, we first study the distributions of $J$ and $J^\dagger$ of each of the baseline clusters. In Figure~\ref{fig:jc_example}, we show two examples of the distributions of $J_{s}^{r, \rm{max}}$ and $J_{s}^{r\dagger, \rm{max}}$ for two baseline clusters. To define stability, we take $50\%$~as the cut in $J_{s}^{r, \rm{max}}$ and $J_{s}^{r\dagger, \rm{max}}$ distributions. When over half of the cluster members in the baseline/random cluster are present in the corresponding random/baseline cluster, we say that the baseline cluster is successfully recovered in that random realization, and thus stable. Although the $50\%$~cut is a choice we make, we verified in the Jaccard distributions of all clusters (similarly to those shown in Figure~\ref{fig:jc_example}) that the stability of the cluster is not sensitive to the exact choice of the cut and is in most cases unambiguous.

More specifically, for a baseline cluster $s$ showing consistently high Jaccard and inverse Jaccard coefficients, specifically $J_s^{r, \rm{max}}> 50\%$ and $J_s^{\dagger r, \rm{max}}>50\%$ for all realizations $r$, the baseline cluster is stable against splitting. Similarly, baseline clusters $s$ with consistently $J_s^{r, \rm{max}}<50\%$ and $J_s^{\dagger r, \rm{max}}<50\%$ are unstable. The more interesting cases are when the $J$ and $J^\dagger$ distributions are inconsistent, as shown in the bottom panel of Figure \ref{fig:jc_example}; the high $J_s$ but low $J_s^\dagger$ distributions indicate that the baseline cluster is either absorbing more members, or merging with other clusters in random realizations. We can distinguish the two cases by examining if multiple baseline clusters share the same maximum $J$ cluster in a given random realization. Lastly, we have baseline clusters with the opposite distributions, low $J$ but high $J^\dagger$; these clusters split in random realizations. Applied to the high-stability sample described earlier in this section, we find that most of the Jaccard distributions across the two clustering spaces show consistently high distributions of $J$ and $J^\dagger$.\footnote{This is not the case if we run \HDBSCAN~over multiple realizations with the original sample, as a large fraction of the clusters merges and splits, making it difficult to identify the robust groups of stars, and their core member stars.} These clusters are identified as the final stable clusters, and the stars within them as core stable stellar members.

\section{Clustering Results}
\label{sec:cluster_res}

We perform \HDBSCAN~for the high-stability sample (defined in Section \ref{sec:stability}) to obtain the clusters from the stable stars. We repeat the process for both clustering spaces (see Section \ref{sec:clustering_spaces}). Note that the high-stability samples are different in each of the clustering spaces (and depend on the \Nstackcut, which we set to 20 as explained in Section \ref{sec:nstackcut}). All clustering summary results shown here are from a single realization with the mean kinematic parameters of the stars. As discussed in Section~\ref{sec:clustering_algorithm}, the clustering procedure is repeated for 100 random realizations, and all clusters are examined for their stability (see Section \ref{sec:stability}). The reported stable clusters in Tables~\ref{tab:vel_cyl_20_stable}~and~\ref{tab:E_act_20_stable} are summarized based on the Jaccard stability studies. Therefore, the clusters from the summary figures may not all be classified as stable in the tables. We find 1405 stars in 23 stable clusters in velocity space, and 497 stars in 6 stable clusters in energy and action space.

\subsection{Velocity clusters}
\label{sec:velocity_clusters}

\begin{figure*}
\begin{center}
\includegraphics[angle=0,width=0.95\textwidth]{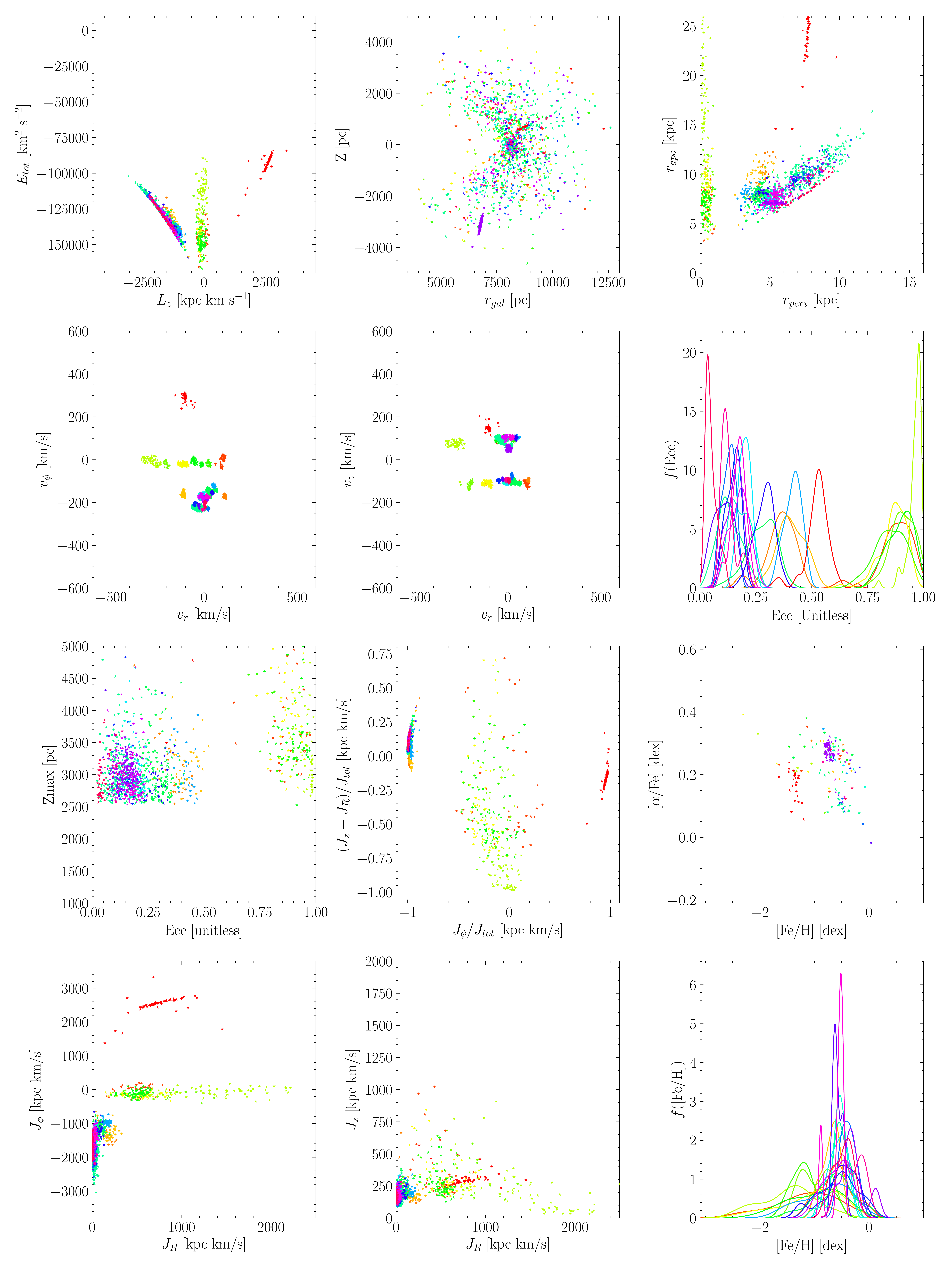}
\end{center}
\caption{Summary plot for clusters that \HDBSCAN~identified from the high-stability (\Nstack$>20$) sample with \zmax~2\,$\sigma$ above 2.5 kpc from the cylindrical velocity space.}
\label{vel_stack_20_summary}
\end{figure*}

Running the above clustering procedure in velocity space, we show the clustering results in Figure~\ref{vel_stack_20_summary}. The stars are color-coded by the clusters with which they are associated. We note that even with this high-stability sample clustering space, not all stars are classified to be in one of the final clusters. The final stable clusters from 100 random realizations of \HDBSCAN~are listed in Table~\ref{tab:vel_cyl_20_stable}.

\makeatletter\onecolumngrid@push\makeatother
\clearpage
\makeatletter\onecolumngrid@pop\makeatother

We see prograde structures with velocity distributions similar to that of the thin and thick disk. One noticeable difference is the slight bias towards positive radial velocity, i.e., more of the identified disk-like prograde clusters are found to be moving away from the galactic center.
We discuss each stable cluster identified here in more details in Section~\ref{sec:discussion}. 

\begin{deluxetable*}{ccccccccccccc}
\tablecaption{Results from sample with \Nstack~greater than 20 in the cylindrical velocity clustering space}
\label{tab:vel_cyl_20_stable}
\tablehead{
\colhead{Cluster ID} & 
\colhead{Color} & 
\colhead{$v_{R}$} & 
\colhead{$\sigma_{v_{R}}$} & 
\colhead{$v_{\phi}$} & 
\colhead{$\sigma_{v_{\phi}}$} & 
\colhead{$v_{z}$} & 
\colhead{$\sigma_{v_{z}}$} & 
\colhead{$ecc$} & 
\colhead{$[Fe/H]$} & 
\colhead{$N_{[Fe/H]}$} & 
\colhead{$N_{star}$} &
\colhead{Stability}
\\
\hline
 \colhead{ } & \colhead{ } & \colhead{$\mathrm{km\,s^{-1}}$} & \colhead{$\mathrm{km\,s^{-1}}$} & \colhead{$\mathrm{km\,s^{-1}}$} & \colhead{$\mathrm{km\,s^{-1}}$} & \colhead{$\mathrm{km\,s^{-1}}$} & \colhead{$\mathrm{km\,s^{-1}}$} & \colhead{ } & \colhead{$\mathrm{dex}$} & \colhead{ } & \colhead{ }
}
\startdata
0 & \textcolor[rgb]{1.0,0.0,0.0}{$\blacksquare$} & -100.93 & 14.21 & 291.85 & 15.01 & 145.05 & 16.55 & 0.53 &  & 0 & 63 & Stable \\
1 & \textcolor[rgb]{1.0,0.2547796665443724,0.0}{$\blacksquare$} & 97.72 & 11.55 & -2.75 & 19.65 & -112.26 & 10.35 & 0.88 & -1.16 & 9 & 39 & Stable \\
2 & \textcolor[rgb]{1.0,0.5095593330887448,0.0}{$\blacksquare$} & 110.0 & 4.39 & -171.9 & 5.72 & -95.88 & 4.74 & 0.36 & -0.61 & 7 & 26 & Stable \\
3 & \textcolor[rgb]{1.0,0.764338999633117,0.0}{$\blacksquare$} & -111.46 & 5.48 & -159.53 & 10.44 & -106.72 & 5.86 & 0.41 & -0.7 & 11 & 42 & Stable \\
4 & \textcolor[rgb]{0.9727937779408368,0.9919124441183264,0.0}{$\blacksquare$} & -109.77 & 16.78 & -21.44 & 7.12 & -110.45 & 9.18 & 0.86 & -1.07 & 16 & 59 & Stable \\
5 & \textcolor[rgb]{0.726101667278138,1.0,0.0}{$\blacksquare$} & -281.08 & 26.82 & -5.67 & 13.93 & 72.8 & 25.23 & 0.96 & -1.47 & 8 & 64 & Stable \\
6 & \textcolor[rgb]{0.4713220007337658,1.0,0.0}{$\blacksquare$} & -204.84 & 12.42 & -19.4 & 11.72 & -117.92 & 13.3 & 0.91 & -0.99 & 2 & 24 & Stable \\
7 & \textcolor[rgb]{0.21654233418939306,1.0,0.0}{$\blacksquare$} & 24.75 & 8.75 & -19.96 & 9.28 & 93.65 & 7.35 & 0.85 & -1.11 & 4 & 26 & Stable \\
8 & \textcolor[rgb]{0.0003669599257835228,1.0,0.061766080148432954}{$\blacksquare$} & -41.78 & 19.17 & -11.6 & 11.1 & 97.05 & 8.69 & 0.89 & -0.99 & 15 & 56 & Stable \\
9 & \textcolor[rgb]{0.0,1.0,0.3161771885991356}{$\blacksquare$} & 36.39 & 14.65 & -151.04 & 11.71 & -106.22 & 6.46 & 0.28 & -0.65 & 26 & 167 & Stable \\
10 & \textcolor[rgb]{0.0,1.0,0.5709552500401159}{$\blacksquare$} & -24.14 & 23.2 & -225.72 & 8.45 & 88.6 & 15.14 & 0.13 & -0.54 & 49 & 251 & Unstable \\
11 & \textcolor[rgb]{0.0,1.0,0.825733311481096}{$\blacksquare$} & -21.08 & 3.87 & -171.0 & 8.7 & -102.21 & 5.74 & 0.18 & -0.55 & 11 & 54 & Stable \\
12 & \textcolor[rgb]{0.0,0.9194881198557668,1.0}{$\blacksquare$} & -2.5 & 3.59 & -160.82 & 5.33 & -111.6 & 4.52 & 0.2 & -0.56 & 10 & 37 & Stable \\
13 & \textcolor[rgb]{0.0,0.6647084533113945,1.0}{$\blacksquare$} & 52.34 & 5.4 & -121.0 & 5.82 & 104.6 & 3.98 & 0.42 & -0.78 & 8 & 34 & Stable \\
14 & \textcolor[rgb]{0.0,0.40992878676702216,1.0}{$\blacksquare$} & 21.54 & 2.52 & -230.59 & 2.77 & -75.43 & 10.34 & 0.13 & -0.38 & 5 & 28 & Unstable \\
15 & \textcolor[rgb]{0.0,0.15514912022264993,1.0}{$\blacksquare$} & -49.2 & 4.82 & -207.99 & 5.06 & -99.18 & 5.09 & 0.15 & -0.54 & 5 & 26 & Unstable \\
16 & \textcolor[rgb]{0.12279233418939263,0.0,1.0}{$\blacksquare$} & 43.19 & 3.43 & -153.73 & 5.16 & 100.18 & 7.57 & 0.29 & -0.55 & 9 & 32 & Stable \\
17 & \textcolor[rgb]{0.3775720007337653,0.0,1.0}{$\blacksquare$} & -27.67 & 3.01 & -214.13 & 6.3 & -90.66 & 4.52 & 0.1 & -0.58 & 3 & 24 & Unstable \\
18 & \textcolor[rgb]{0.632351667278138,0.0,1.0}{$\blacksquare$} & 5.53 & 6.88 & -183.34 & 7.56 & 51.14 & 7.83 & 0.15 & -0.3 & 6 & 184 & Stable \\
19 & \textcolor[rgb]{0.8871313338225106,0.0,1.0}{$\blacksquare$} & -15.19 & 8.52 & -172.05 & 8.28 & 103.42 & 7.02 & 0.17 & -0.59 & 16 & 63 & Stable \\
20 & \textcolor[rgb]{1.0,0.0,0.858088999633117}{$\blacksquare$} & 18.78 & 7.3 & -172.49 & 4.68 & 104.28 & 5.6 & 0.17 & -0.59 & 5 & 39 & Ambiguous \\
21 & \textcolor[rgb]{1.0,0.0,0.6033093330887447}{$\blacksquare$} & -7.8 & 4.65 & -233.56 & 2.34 & -94.81 & 6.03 & 0.13 & -0.31 & 2 & 20 & Unstable \\
22 & \textcolor[rgb]{1.0,0.0,0.348529666544373}{$\blacksquare$} & 4.87 & 3.06 & -208.89 & 8.21 & -98.67 & 7.19 & 0.05 & -0.48 & 13 & 47 & Unstable \\
\enddata
\end{deluxetable*}

\subsection{Integrals of Motion Clusters}
\label{sec:action_clusters}

\begin{figure*}
\begin{center}
\includegraphics[angle=0,width=0.95\textwidth]{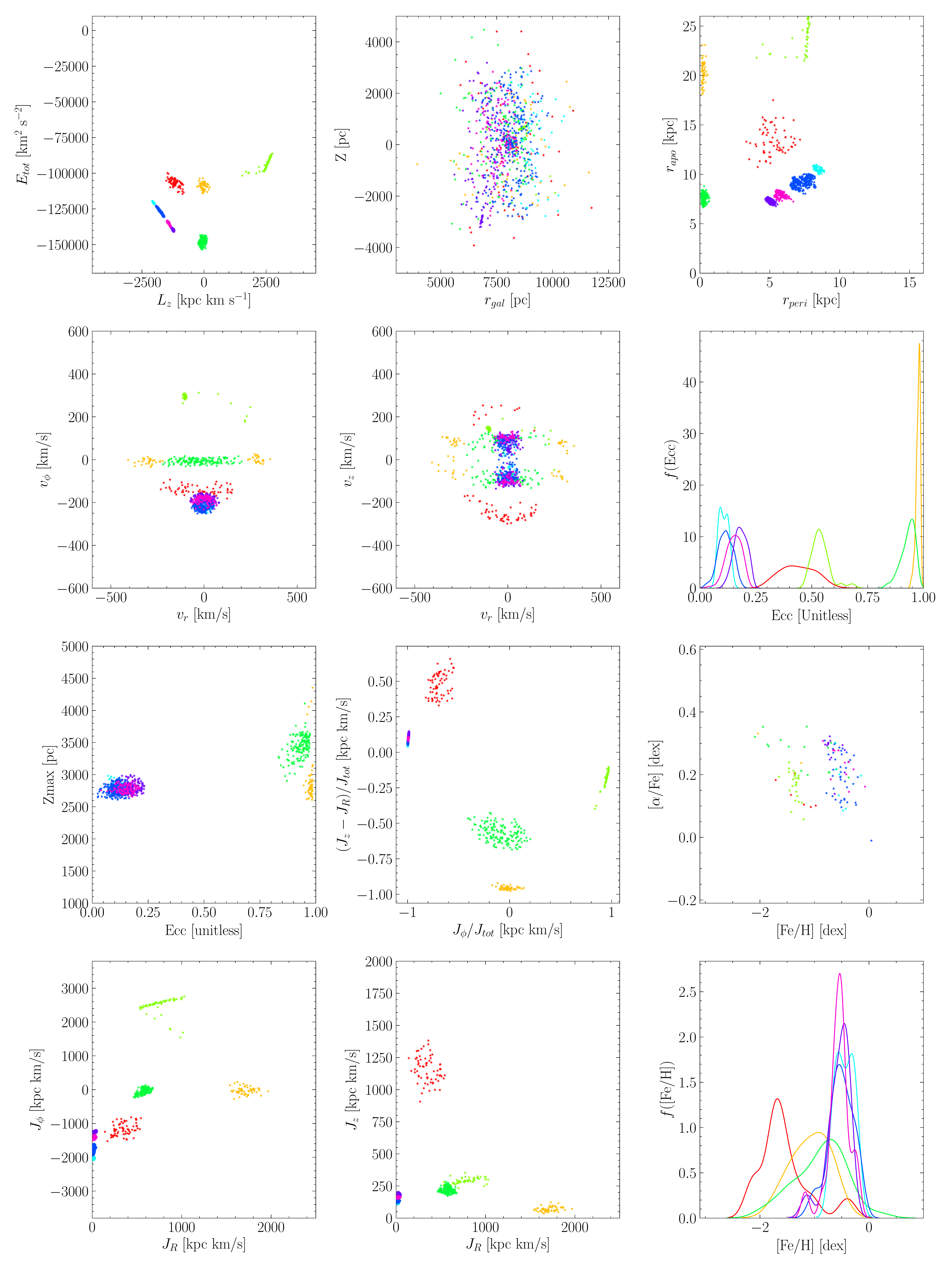}
\end{center}
\caption{Summary plot for clusters that \HDBSCAN~identified from the high-stability (\Nstack$>20$) sample with \zmax~2\,$\sigma$ above 2.5 kpc from the integral of motion space.}
\label{Eact_stack_20_summary}
\end{figure*}

The clustering results in the Integrals of Motion (Energy and Action) space applied to the high-stability sample are shown in Figure~\ref{Eact_stack_20_summary}. As expected, the clustering results are not the same as those identified in the cylindrical velocity space, as seen in Figure~\ref{vel_stack_20_summary}. We examine and summarize the properties and stabilities of these clusters in Table~\ref{tab:E_act_20_stable}. 

\makeatletter\onecolumngrid@push\makeatother
\begin{rotatetable*}
\begin{deluxetable*}{ccccccccccccccc}
\centerwidetable
\tablecaption{Results from sample with \Nstack~greater than 20 in the energy and action clustering space}
\label{tab:E_act_20_stable}
\tabletypesize{\footnotesize}
\tablehead{
\colhead{Cluster ID} & 
\colhead{Color} & 
\colhead{$J_{R}$} & 
\colhead{$\sigma_{J_{R}}$} & 
\colhead{$J_{\phi}$} & 
\colhead{$\sigma_{J_{\phi}}$} & 
\colhead{$J_{z}$} & 
\colhead{$\sigma_{J_{z}}$} & 
\colhead{$E_{tot}$} & 
\colhead{$\sigma_{E_{tot}}$} & 
\colhead{$ecc$} & 
\colhead{$[Fe/H]$} & 
\colhead{$N_{[Fe/H]}$} & 
\colhead{$N_{star}$} &
\colhead{Stability}
\\ \hline
\colhead{ } & \colhead{ } & \colhead{$\mathrm{km\,kpc\,s^{-1}}$} & \colhead{$\mathrm{km\,kpc\,s^{-1}}$} & \colhead{$\mathrm{km\,kpc\,s^{-1}}$} & \colhead{$\mathrm{km\,kpc\,s^{-1}}$} & \colhead{$\mathrm{km\,kpc\,s^{-1}}$} & \colhead{$\mathrm{km\,kpc\,s^{-1}}$} & \colhead{$\mathrm{km^{2}\,s^{-2}}$} & \colhead{$\mathrm{km^{2}\,s^{-2}}$} & \colhead{ } & \colhead{$\mathrm{dex}$} & \colhead{ } & \colhead{ } & \colhead{ }
}
\startdata
0 & \textcolor[rgb]{1.0,0.0,0.0}{$\blacksquare$} & 349.28 & 97.08 & -1167.43 & 178.0 & 1155.14 & 99.66 & -106802.56 & 2941.69 & 0.44 & -1.56 & 12 & 65 & Stable \\
1 & \textcolor[rgb]{1.0,0.741177211765447,0.0}{$\blacksquare$} & 1713.61 & 90.77 & -27.99 & 104.25 & 69.67 & 17.76 & -109012.71 & 2403.22 & 0.98 & -1.07 & 7 & 55 & Stable \\
2 & \textcolor[rgb]{0.5176455764691059,1.0,0.0}{$\blacksquare$} & 757.92 & 134.15 & 2486.99 & 242.59 & 286.34 & 25.2 & -93541.68 & 3778.38 & 0.54 &  & 0 & 59 & Stable \\
3 & \textcolor[rgb]{0.0,1.0,0.22353062080241548}{$\blacksquare$} & 573.94 & 47.09 & -52.75 & 82.0 & 217.45 & 21.46 & -147907.82 & 2284.24 & 0.93 & -0.89 & 34 & 155 & Stable \\
4 & \textcolor[rgb]{0.0,1.0,0.9647031631761764}{$\blacksquare$} & 16.38 & 6.44 & -2033.83 & 23.38 & 119.05 & 5.36 & -120579.56 & 549.61 & 0.11 & -0.45 & 16 & 57 & Stable \\
5 & \textcolor[rgb]{0.0,0.29411984742867114,1.0}{$\blacksquare$} & 17.11 & 9.01 & -1760.66 & 86.17 & 133.7 & 7.04 & -126802.74 & 2026.59 & 0.11 & -0.51 & 58 & 287 & Unstable \\
6 & \textcolor[rgb]{0.44705736433677606,0.0,1.0}{$\blacksquare$} & 31.23 & 10.13 & -1248.4 & 25.81 & 180.78 & 9.29 & -139561.44 & 641.08 & 0.18 & -0.56 & 14 & 131 & Unstable \\
7 & \textcolor[rgb]{1.0,0.0,0.8117654238977766}{$\blacksquare$} & 24.86 & 10.26 & -1422.58 & 41.28 & 161.64 & 4.6 & -135038.12 & 1151.29 & 0.15 & -0.56 & 19 & 106 & Stable \\
\enddata
\end{deluxetable*}
\end{rotatetable*}
\clearpage
\makeatletter\onecolumngrid@pop\makeatother

\makeatletter\onecolumngrid@push\makeatother
\clearpage
\makeatletter\onecolumngrid@pop\makeatother

\subsection{Relationship between clusters} 
\label{sec:cluster_relation}

\begin{figure*}
\begin{center}
\includegraphics[angle=0,width=0.90\textwidth]{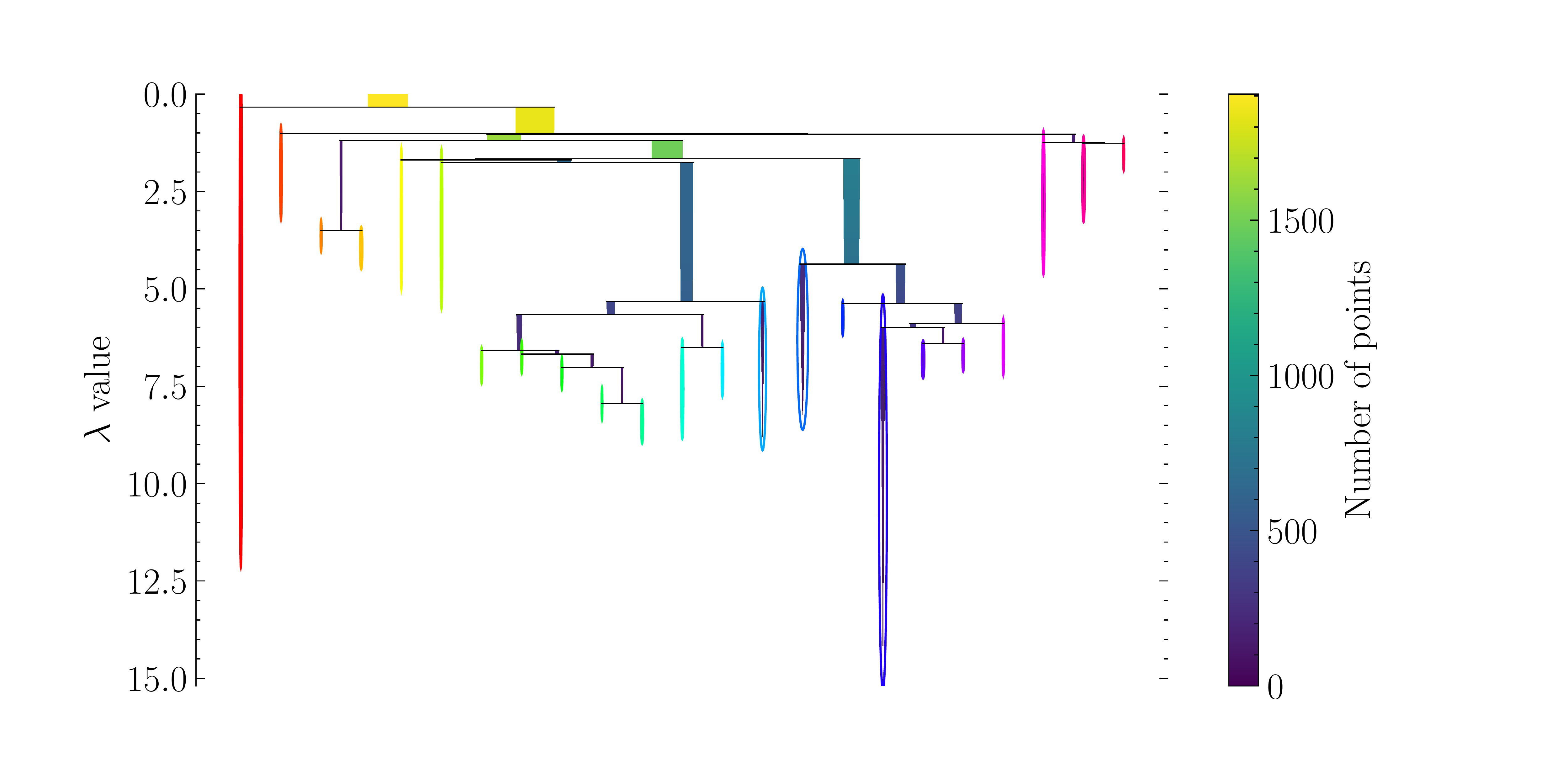}
\end{center}
\caption{Condensed merger tree for the stacked subsample with \Nstack$>20$ and \zmax~2\,$\sigma$ above 2.5 kpc from the cylindrical velocity space. The vertical axis is $\lambda$, a measure of distance that \HDBSCAN~uses to define the edges of the cluster. See \cite{mcinnes2017hdbscan} for more details.}
\label{vel_stacked_20_2p5_tree}
\end{figure*}

We examine the \HDBSCAN~merger trees from the clustering results mentioned in Sections~\ref{sec:velocity_clusters} and \ref{sec:action_clusters}. The trees are extracted by the \texttt{condensed\_tree\_}~method from the \HDBSCAN~package. The merger tree allows us to directly see how clusters are related to each other along the hierarchical tree. This is especially important for our clustering method, \texttt{leaf}, which captures the smallest clusters in the sample and is thus designed to subdivide clusters that may be related to each other. It also provides intuition on which axis \HDBSCAN~primarily uses to distinguish clusters during the clustering process. Figure~\ref{vel_stacked_20_2p5_tree} shows the merger tree for the clustering in velocity space. Identified clusters are shown with a color matching that given in Figure~\ref{vel_stack_20_summary}. By examining the Figure, we find, for example, that Cluster 0 from Table~\ref{tab:vel_cyl_20_stable} is the first to be separated from the rest of the clusters, which is sensible given that Cluster 0 is the only retrograde structure found, and it is surrounded by low density points that \HDBSCAN~deems to be noise. The stars are generally first separated by \vphi, where we see a clear division between clusters with and without prograde kinematics.

To study whether multiple clusters belong to the same object, we perform a KS test \citep{kolmogorov33,smirnov48} on the available metallicity of each cluster, which is obtained through cross-matching from LAMOST DR6 \citep{cui12,zhao12}.\footnote{We note that we use the metallicity measurements from LAMOST DR6 for better statistics in the comparison, as they provide more metallicities than APOGEE DR17, which were featured in the summary plots.} 
Note that while we classified the clusters' stabilities based on their $J$ and $J^\dagger$ distributions, all clusters from the baseline realization, stable and unstable, are included in the following analysis to avoid inconsistencies caused by removing nodes in the merger tree. Some of the instability may be reflected during the merging process, so it is useful to keep them while inspecting the final results. We note that due to the lack of metallicity measurements for part of the sample, some of the smaller clusters contain too few valid metallicity measurements to provide enough statistics to perform a KS test. In cases where there are fewer than five available [Fe/H] measurements in a given cluster, we skip the comparison, move on to the parent node, and assume the child node can be merged.

We work from the bottom up to compare the metallicities of the clusters on each of the branches. If they are statistically consistent with being drawn from the same group, we merge them together. If not, we keep them as separate groups. We use the \texttt{KS\_2samp} routine from the \texttt{Scipy} package \citep{jones01} to perform the KS test. The results are discussed further in Section~\ref{sec:discussion}.

\makeatletter\onecolumngrid@push\makeatother
\clearpage
\makeatletter\onecolumngrid@pop\makeatother

\section{Discussion and Interpretation}
\label{sec:discussion}

We compare our final clustering results in the two clustering spaces (see Section \ref{sec:clustering_spaces}), with parameters set at \zmax~$>2.5$ kpc, and \Nstack~$>20$, against known Milky Way substructures from the literature. 

\subsection{Merged clusters}

\begin{figure*}
\begin{center}
\includegraphics[angle=0,width=0.45\textwidth]{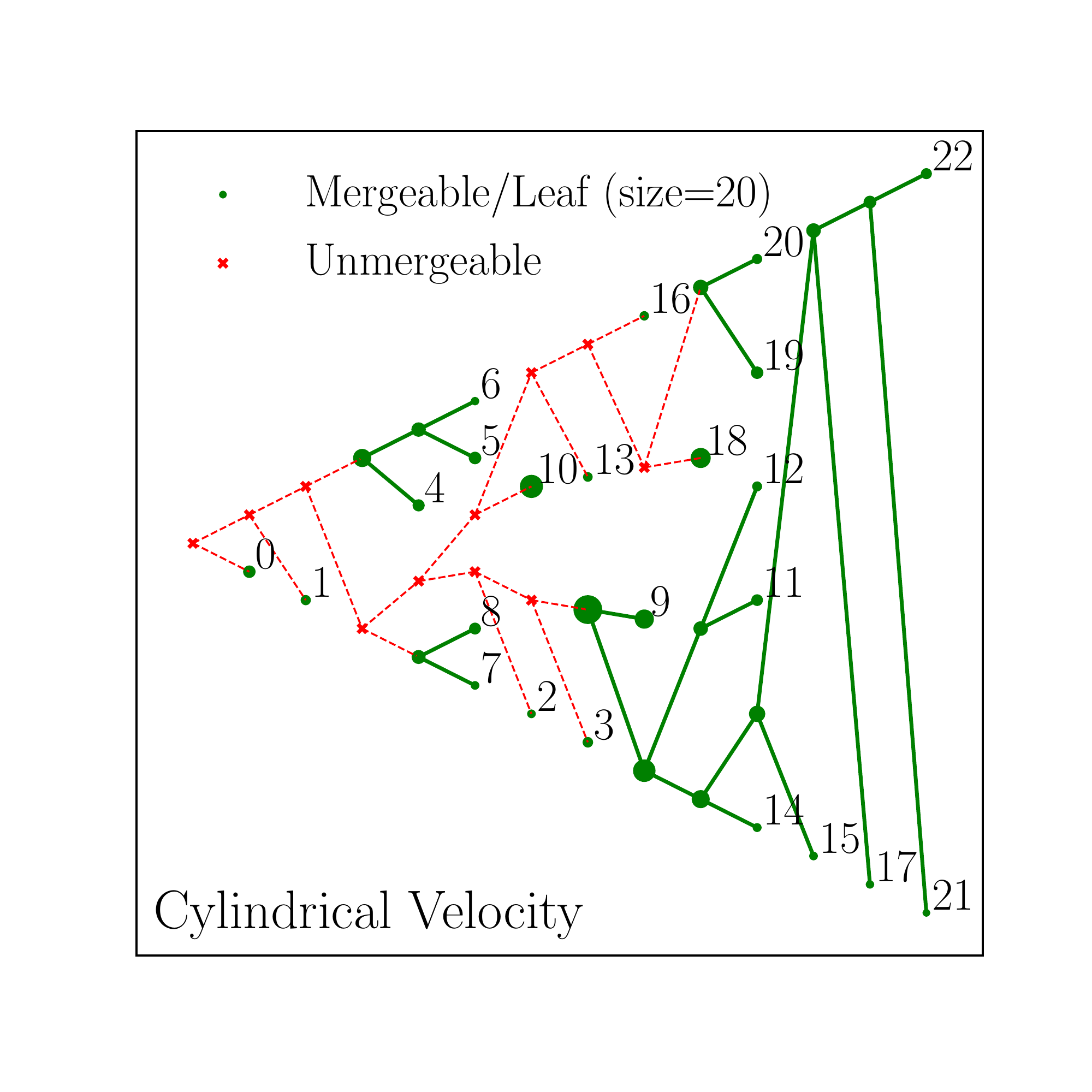} 
\includegraphics[angle=0,width=0.45\textwidth]{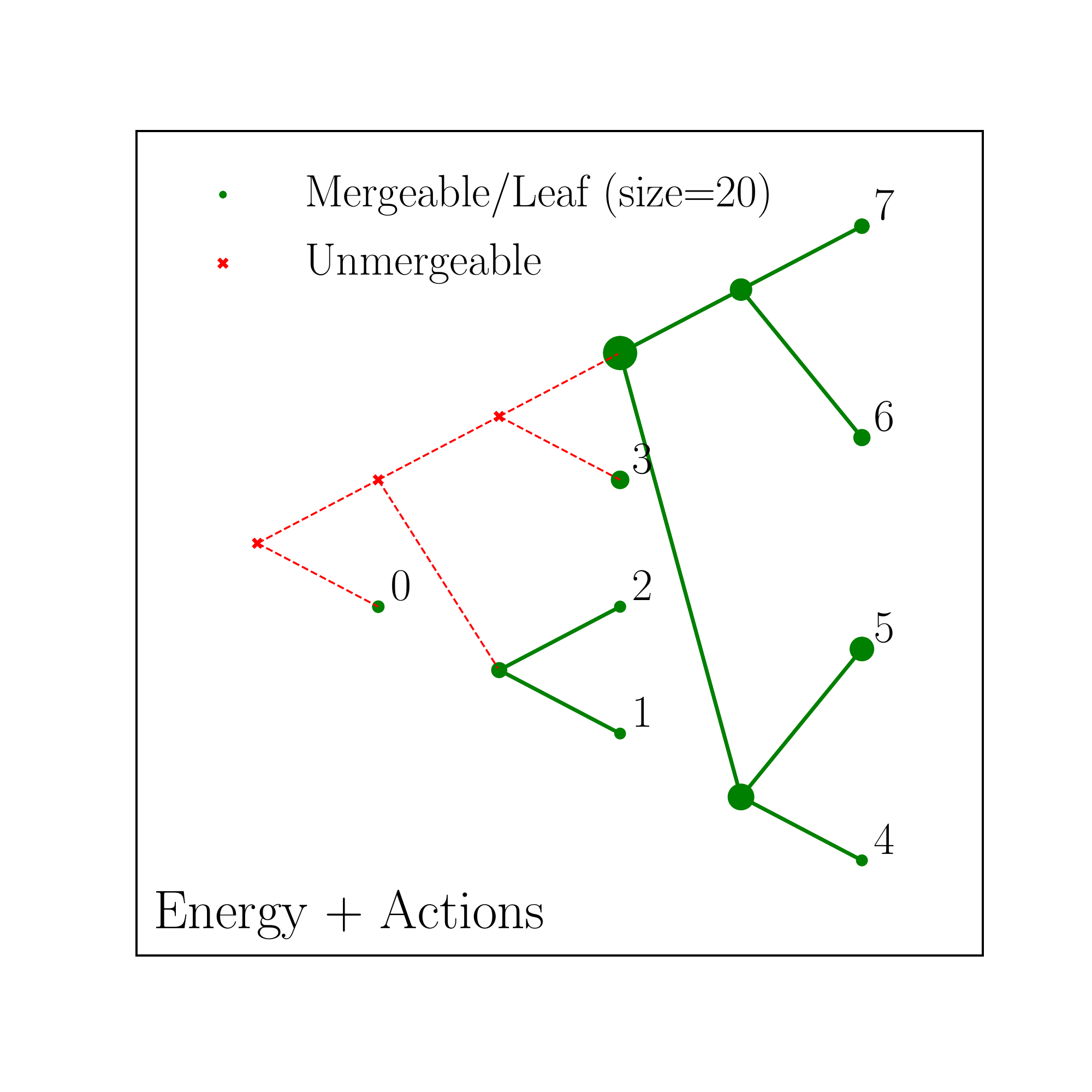}
\end{center}
\caption{Semi-tree diagrams showing the nodes for the stacked stable subsample with \zmax~2\,$\sigma$ above 2.5 kpc from the cylindrical velocity space (left) and the energy and action space (right). Size of the nodes is proportional to the size of cluster after merging children nodes. Green in general means the clusters can be merged together either because one or both of the children nodes have less than 10 member or because the they passed the metallicity KS test. Red means the node could not be merged either because the test mentioned above failed or one of the children nodes failed to merge.
}
\label{vel_E_stacked_20_2p5_merge}
\end{figure*}

From Figure~\ref{vel_E_stacked_20_2p5_merge} (left panel) and Tables~\ref{tab:vel_cyl_20_stable} and \ref{tab:vel_cyl_20_stable_merge}, we see that the unstable clusters 14, 15, 17, 21, and 22 are all mergeable (along with the stable clusters 9, 11, and 12). This indicates that the instability of these five clusters against splitting and merging is consistent with the fact that they are compatible with being part of a larger cluster in metallicity space. They are likely all coming from the same larger cluster in velocity space, but only split/merge due to their locations in phase space and the associated uncertainties. Thus, these unstable clusters are fragments of a larger cluster, split in different ways. The member stars in these clusters still remain robust, albeit for a larger cluster.

On the other hand, the unstable cluster 10 has no potentially merging clusters in metallicity space, suggesting that cluster 10 by itself is unlikely a robust cluster and dissolves in velocity space when considering uncertainties. Stellar members in this cluster are thus not robust cluster members.

In addition to the previously identified unstable clusters, we also get stable clusters merged, which is the case for clusters 4-9, 11, 12, 19, and 20 from Table~\ref{tab:vel_cyl_20_stable}. Their corresponding parents, after merging, provide hints for the reason. For example, the newly merged clusters \RN{6} and \RN{10} are likely both part of the thick disk, given the prograde rotational velocity of \vphi~at $\sim -180$ km\,${\text{s}^{-1}}$. The \texttt{leaf} method we used for \HDBSCAN~divided them into smaller subgroups but their metallicities distributions allow us to confidently say that they can be merged back into a larger group. 

\begin{figure*}
\begin{center}
\includegraphics[angle=0,width=0.95\textwidth]{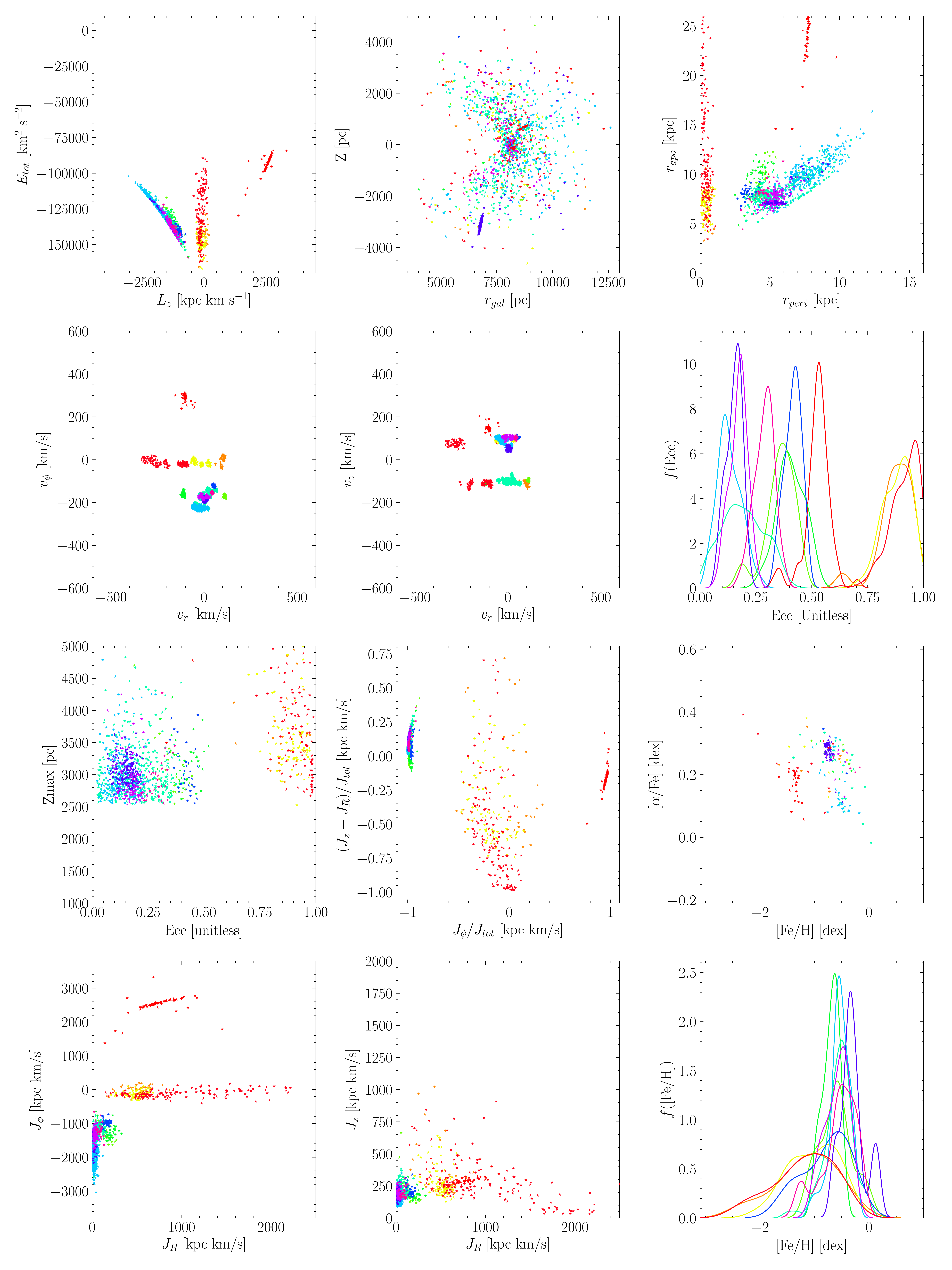}
\end{center}
\caption{Summary plot for merged clusters that \HDBSCAN~identified in the high-stability (\Nstack$>20$) sample with \zmax~2\,$\sigma$ above 2.5 kpc from the cylindrical velocity space. Note the colors do not match those from Figure~\ref{vel_stack_20_summary}, but those quoted in Table~\ref{tab:vel_cyl_20_stable_merge}.}
\label{vel_stack_20_merged_summary}
\end{figure*}

\begin{figure*}
\begin{center}
\includegraphics[angle=0,width=0.95\textwidth]{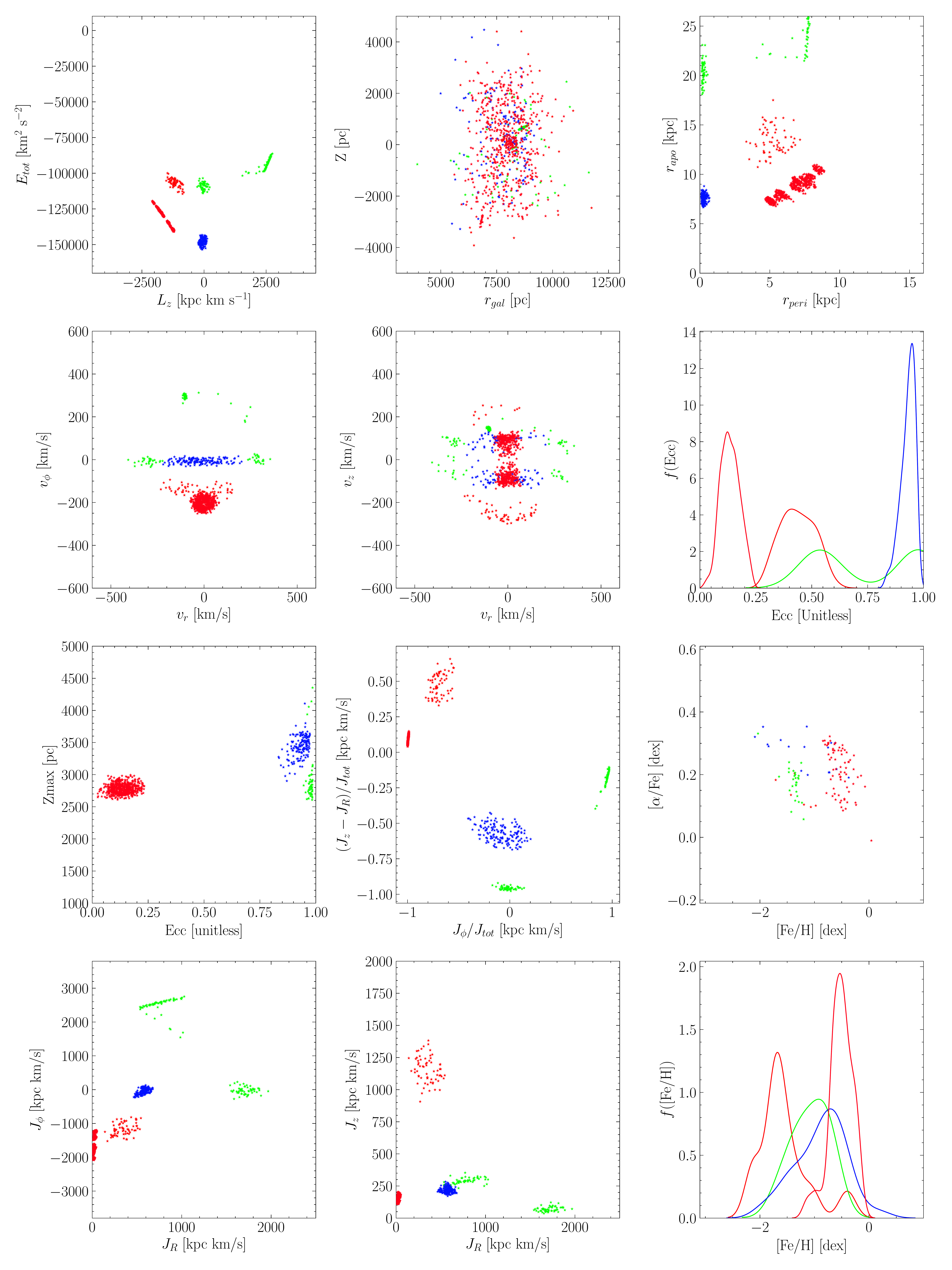}
\end{center}
\caption{Summary plot for merged clusters that \HDBSCAN~identified in the high-stability (\Nstack$>20$) sample with \zmax~2\,$\sigma$ above 2.5 kpc from the energy and action space. Note the colors do not match those from Figure~\ref{Eact_stack_20_summary}}
\label{Eact_stack_20_merged_summary}
\end{figure*}

We repeat the same process with the clustering result for the integrals of motion space and include the results in Figure~\ref{vel_E_stacked_20_2p5_merge} (right panel) and Table~\ref{tab:E_act_20_stable_merge}. Clusters 5 and 6, which were previously identified as unstable in Table~\ref{tab:E_act_20_stable}, are merged together with clusters 4 and 7. These results are shown in Figures~\ref{vel_stack_20_merged_summary} and \ref{Eact_stack_20_merged_summary} for the velocity and integral of motion spaces, respectively. 
We discuss in the following subsections how the clusters compare with structures known from previous studies. When applicable, we use references from the literature where the kinematic values are obtained and derived similarly to our study to avoid systematic differences, i.e., original data from \Gaia and similar galaxy potentials used in orbital integrations.

\makeatletter\onecolumngrid@push\makeatother
\clearpage
\makeatletter\onecolumngrid@pop\makeatother

\makeatletter\onecolumngrid@push\makeatother
\begin{rotatetable*}
\begin{deluxetable*}{ccccccccccccc}
\centerwidetable
\tablecaption{Clusters merged based on metallicity distribution from Tab~\ref{tab:vel_cyl_20_stable}.}
\label{tab:vel_cyl_20_stable_merge}
\tablehead{
\colhead{Cluster ID} & 
\colhead{Color} &
\colhead{$v_{R}$} & 
\colhead{$\sigma_{v_{R}}$} & 
\colhead{$v_{\phi}$} & 
\colhead{$\sigma_{v_{\phi}}$} & 
\colhead{$v_{z}$} & 
\colhead{$\sigma_{v_{z}}$} & 
\colhead{$ecc$} & 
\colhead{$[Fe/H]$} & 
\colhead{$N_{[Fe/H]}$} & 
\colhead{$N_{star}$} & 
\colhead{Merged}
\\ \hline
\colhead{ } & \colhead{ } & \colhead{$\mathrm{km\,s^{-1}}$} & \colhead{$\mathrm{km\,s^{-1}}$} & \colhead{$\mathrm{km\,s^{-1}}$} & \colhead{$\mathrm{km\,s^{-1}}$} & \colhead{$\mathrm{km\,s^{-1}}$} & \colhead{$\mathrm{km\,s^{-1}}$} & \colhead{ } & \colhead{$\mathrm{dex}$} & \colhead{ } & \colhead{ } & \colhead{ }
}
\startdata
\RN{1} & \textcolor[rgb]{1.0,0.0,0.0}{$\blacksquare$} & -100.93 & 14.21 & 291.85 & 15.01 & 145.05 & 16.55 & 0.53 & \nodata & 0 & 63 & [0] \\
\RN{2} & \textcolor[rgb]{1.0,0.532721120956415,0.0}{$\blacksquare$} & 97.72 & 11.55 & -2.75 & 19.65 & -112.26 & 10.35 & 0.88 & -1.16 & 9 & 39 & [1] \\
\RN{3} & \textcolor[rgb]{0.93455775808717,1.0,0.0}{$\blacksquare$} & -20.69 & 35.12 & -14.25 & 11.25 & 95.97 & 8.44 & 0.88 & -1.02 & 19 & 82 & [7 8] \\
\RN{4} & \textcolor[rgb]{0.4018366371307548,1.0,0.0}{$\blacksquare$} & 110.0 & 4.39 & -171.9 & 5.72 & -95.88 & 4.74 & 0.36 & -0.61 & 7 & 26 & [2] \\
\RN{5} & \textcolor[rgb]{0.0,1.0,0.15404569495487538}{$\blacksquare$} & -111.46 & 5.48 & -159.53 & 10.44 & -106.72 & 5.86 & 0.41 & -0.7 & 11 & 42 & [3] \\
\RN{6} & \textcolor[rgb]{0.0,1.0,0.6867634597860162}{$\blacksquare$} & 8.88 & 29.44 & -178.41 & 31.77 & -101.21 & 10.86 & 0.19 & -0.56 & 75 & 403 & [9 11 12 14 15 17 21 22] \\
\RN{7} & \textcolor[rgb]{0.0,0.7805173926497455,1.0}{$\blacksquare$} & -24.14 & 23.2 & -225.72 & 8.45 & 88.6 & 15.14 & 0.13 & -0.54 & 49 & 251 & [10] \\
\RN{8} & \textcolor[rgb]{0.0,0.24779627169333063,1.0}{$\blacksquare$} & 52.34 & 5.4 & -121.0 & 5.82 & 104.6 & 3.98 & 0.42 & -0.78 & 8 & 34 & [13] \\
\RN{9} & \textcolor[rgb]{0.30808663713075457,0.0,1.0}{$\blacksquare$} & 5.53 & 6.88 & -183.34 & 7.56 & 51.14 & 7.83 & 0.15 & -0.3 & 6 & 184 & [18] \\
\RN{10} & \textcolor[rgb]{0.8408077580871701,0.0,1.0}{$\blacksquare$} & -2.2 & 18.38 & -172.21 & 7.13 & 103.75 & 6.53 & 0.17 & -0.59 & 21 & 102 & [19 20] \\
\RN{11} & \textcolor[rgb]{1.0,0.0,0.6264711209564149}{$\blacksquare$} & 43.19 & 3.43 & -153.73 & 5.16 & 100.18 & 7.57 & 0.29 & -0.55 & 9 & 32 & [16] \\
\RN{12} & \textcolor[rgb]{1.0,0.0,0.09375}{$\blacksquare$} & -199.88 & 81.15 & -14.24 & 13.58 & -31.89 & 93.79 & 0.91 & -1.19 & 26 & 147 & [4 5 6] \\
\enddata
\end{deluxetable*}
\end{rotatetable*}

\clearpage
\makeatletter\onecolumngrid@pop\makeatother
\makeatletter\onecolumngrid@push\makeatother
\begin{rotatetable*}
\begin{deluxetable*}{ccccccccccccccc}
\centerwidetable
\tablecaption{Clusters merged based on metallicity distribution from Tab~\ref{tab:E_act_20_stable}. }
\label{tab:E_act_20_stable_merge}
\tablehead{
\colhead{Cluster ID} & 
\colhead{Color} &
\colhead{$J_{R}$} & 
\colhead{$\sigma_{J_{R}}$} & 
\colhead{$J_{\phi}$} & 
\colhead{$\sigma_{J_{\phi}}$} & 
\colhead{$J_{z}$} & 
\colhead{$\sigma_{J_{z}}$} & 
\colhead{$E_{tot}$} & 
\colhead{$\sigma_{E_{tot}}$} & 
\colhead{$ecc$} & 
\colhead{$[Fe/H]$} & 
\colhead{$N_{[Fe/H]}$} & 
\colhead{$N_{star}$} & 
\colhead{Merged}
\\ 
\hline
\colhead{ } & \colhead{ } & \colhead{$\mathrm{km\,kpc\,s^{-1}}$} & \colhead{$\mathrm{km\,kpc\,s^{-1}}$} & \colhead{$\mathrm{km\,kpc\,s^{-1}}$} & \colhead{$\mathrm{km\,kpc\,s^{-1}}$} & \colhead{$\mathrm{km\,kpc\,s^{-1}}$} & \colhead{$\mathrm{km\,kpc\,s^{-1}}$} & \colhead{$\mathrm{km^{2}\,s^{-2}}$} & \colhead{$\mathrm{km^{2}\,s^{-2}}$} & \colhead{$\mathrm{}$} & \colhead{$\mathrm{dex}$} & \colhead{ } & \colhead{ } & \colhead{ }
}
\startdata
\RN{1} & \textcolor[rgb]{1.0,0.0,0.0}{$\blacksquare$} & 349.28 & 97.08 & -1167.43 & 178.0 & 1155.14 & 99.66 & -106802.56 & 2941.69 & 0.44 & -1.56 & 12 & 65 & [0] \\
\RN{2} & \textcolor[rgb]{0.03124934374934376,1.0,0.0000013125013124790507}{$\blacksquare$} & 1219.0 & 491.27 & 1273.62 & 1270.84 & 181.81 & 110.47 & -101005.77 & 8362.99 & 0.75 & -1.07 & 7 & 114 & [1 2] \\
\RN{3} & \textcolor[rgb]{0.0,0.062501968751969,1.0}{$\blacksquare$} & 573.94 & 47.09 & -52.75 & 82.0 & 217.45 & 21.46 & -147907.82 & 2284.24 & 0.93 & -0.89 & 34 & 155 & [3] \\
\RN{4} & \textcolor[rgb]{1.0,0.0,0.09375}{$\blacksquare$} & 21.64 & 11.04 & -1610.28 & 262.5 & 147.97 & 22.59 & -130571.45 & 6408.31 & 0.14 & -0.52 & 107 & 581 & [4 5 6 7] \\
\enddata
\end{deluxetable*}
\end{rotatetable*}

\clearpage
\makeatletter\onecolumngrid@pop\makeatother

\makeatletter\onecolumngrid@push\makeatother
\clearpage
\makeatletter\onecolumngrid@pop\makeatother

\makeatletter\onecolumngrid@push\makeatother
\clearpage
\makeatletter\onecolumngrid@pop\makeatother

\subsection{Velocity space}

In this section, we associate clusters found in the velocity clustering space with known structures in the literature by discussing their mean kinematics and chemical characteristics in turn. A summary of the all the clusters discussed here can be found in Figure~\ref{vel_stack_20_merged_summary}, and Table~\ref{tab:vel_cyl_20_stable_merge}. 

We start with the two globular clusters identified, NGC 3201 (cluster \RN{1} shown as dark red in Figure~\ref{vel_stack_20_merged_summary}) and NGC 104 (cluster \RN{9}, shown as dark blue in Figure~\ref{vel_stack_20_merged_summary}). We find these clusters at $(l, b) = (277.21,8.63)$ degrees and $(305.88,-44.90)$ degrees, and heliocentric distances of 4.22 kpc and 4.33 kpc, respectively. These coordinates are consistent with the values of  $(l, b) = (277.23,8.64)$ degrees and $(305.89,-44.90)$ degrees from \citet{harris96,harris10} considering the half-light radius of these two clusters are at 3.1 arcmin and 3.17 arcmin, respectively. The heliocentric distances in \citet{harris96,harris10} are 4.9 and 4.5 kpc for NGC~3201 and NGC~104, respectively, where we attribute the disagreement with our values to systematic uncertainties in the different methods used in measuring distances, i.e., photometry vs. astrometry. 
Finding these globular clusters within our framework is a validation of our clustering method. It is, however, worth noting that these two globular clusters appear more extended along the line of sight than they actually are from dedicated globular cluster studies (see \citealt{harris10}, and references therein). This is a common effect in galactic coordinate transformation caused by the uncertainties in the distance to the sources. First, this further emphasizes the importance of incorporating uncertainties into the clustering. Additionally, a similar effect would be present for the entire sample and as a result systematically increase the scale height of the thin disk at high galactic latitude. This plays a role in interpreting the prograde structures we identify, which we discuss below. 

Our newly identified clusters \RN{2}, \RN{3}, \RN{12} have almost no rotational motion and low metallicity. Their chemical characteristics are overall consistent with being part of GSE, with mean [Fe/H] = $-$1.16, $-$1.02, and $-$1.19, respectively. These values agree with literature estimates of the metallicity of the GSE, which ranges from [Fe/H]~$\sim -1.6$ to $-1.1$, depending on the selection criterion \citep{helmi18,necib19a,naidu20}. Because of the \zmax~cut we applied, the stars identified as part of these clusters most likely correspond to the high vertical motion portion of the GSE. The majority of the GSE would be left out in the initial cut, yet the fact that the overdensity is still identified suggests the GSE has a non-negligible tail in high vertical velocity distribution. Clusters \RN{2} and \RN{12} (shown in orange and red in Figure~\ref{vel_stack_20_merged_summary}) have the known high radial velocities of the GSE \citep{belokurov18,helmi18}. However, Cluster \RN{3} (shown in yellow) does not. It could be that this cluster, which has eccentricities and metallicities consistent with clusters \RN{2} and \RN{12}, encompasses the stars that connect the positive and negative radial velocity lobes of the GSE.  

We now discuss the possibility of our prograde clusters being the high $J_z$ tail of the thin disk or thick disk. Knowing that the uncertainties in the distances can systematically increase the scale height of the disk, stars originally on the exponential tail of the disk height distribution could be pushed outward into our selection region, \zmax~greater than 2.5 kpc, as was shown earlier in Figure~\ref{zmax_int_summary}. We thus expect there to be a significant amount of disk stars in the final high-stability sample. Cluster \RN{7} (in cyan), despite the significant vertical motion, does otherwise appear to be part of the thin disk. The cluster has rotational velocity $v_{\phi} = -225.72$ km\,${\text{s}^{-1}}$ and eccentricity of 0.13, consistent with being part of the thin disk. The mean metallicity of $-$0.54 and low $\alpha$-abundance also support the thin disk origin.

Additionally, clusters \RN{4}, \RN{5}, \RN{6}, \RN{10}, colored as fluorescent green, green, turquoise, and purple, respectively, sit in between the two groups mentioned above, with kinematics more consistent with that of the thick disk, and average \vphi~approximately 150 to 180 km\,${\text{s}^{-1}}$. The current estimate of the canonical thick disk scale height from recent literature study is $0.9 \pm 0.2$ kpc (see \citet{bland-hawthorn16} for a review on galactic disc parameters). The metallicity distributions of these clusters peak between $-$1 and 0 dex, with overall high $\alpha$-abundances, which are consistent with these stars being part of the canonical thick disk. We thus associate them as part of the thick disk. Clusters \RN{8} (in dark blue) and \RN{11} (in pink), while being prograde, appear distinct from the thin disk with metallicity distributions extending lower than $-$1. This can be taken as a sign of these clusters containing accreted stellar populations or a portion of the metal-weak Atari disk \citep{Norris85,Morrison90,chiba00,mardini22}.

\subsection{Integrals of Motion Space}
\label{sec:integrals}

Similarly to the above, we associate the following clusters identified in energy and action space with known Milky Way substructures. We find four clusters that we list in Table~\ref{tab:E_act_20_stable_merge}, and show in Figure~\ref{Eact_stack_20_merged_summary}.
Cluster \RN{1}, shown in the Figure colored in dark red, closely resembles the Helmi stream \citep{helmi99b}, which is expected due to its well-known large vertical motion. Thus most of its members pass the \zmax~cut. The mean kinematics and metallicity of the cluster identified in our study are $J_{\phi} = -1.17^{\pm 0.10} \times 10^3~\text{kpc\,km\,s}^{-1}$, $E_{tot} = -10.68^{\pm 0.29} \times 10^{4}~\text{km}^2\,\text{s}^{-2}$, and [Fe/H]~$=-1.56$. They agree with literature selection of the Helmi stream in $L_z$ in $[-1.7,-0.75] \times 10^3~\text{kpc\,km\,s}^{-1}$ \citep{koppelman19b}, $E_{tot} = -10.3 \times 10^{4}~\text{km}^2\,\text{s}^{-2}$, and peak [Fe/H] ranging from $-$1.5 to $-$1.3 \citep{naidu20}.  As the merger event associated with these streams happened approximately 5-8 Gyr ago \citep{koppelman19b}, the stream does not show a significant coherent spatial structure. Its stars distribute spatially similar to the background stars, except being confined by Galactic radial distance \citep{koppelman19b}. 

Cluster \RN{2}, shown in green, is a mix of three substructures. Merging based on metallicity alone has caused structures that are obviously not from the same group to be merged. Thus, instead of discussing the merged cluster \RN{2}, we discuss pre-merging clusters 1 and 2, together with cluster \RN{3}, i.e., cluster 3.

The radial motion dominated clusters (pre-merging clusters 1 and 3), resembling the GSE, are identified with much fewer clusters than in velocity space, and they are mainly distinguished by energy rather than radial velocity. Both clusters have mean metallicity [Fe/H]~$\sim -1$, roughly consistent with literature values, but the lower energy cluster 3 shows potential for contamination from the high metallicity in-situ halo at [Fe/H]~$\sim -0.5$ and the big splash (stars formed in protodisc and have orbit altered to the highly eccentric orbit by the last major merger event) with metallicity $>-0.7$ from \citet{belokurov20}. The higher energy cluster, on the other hand, appears to be more in line with the metal-poor nature of an accreted structure like the GSE. 

Globular cluster NGC 3201 (our pre-merging cluster 2) is identified as in the cylindrical velocity space. There are, however, a few additional members identified in this cluster, which are most likely not true members of the globular cluster. We suspect they actually belong to the high energy retrograde structure Sequoia \cite{myeong19}, and are instead mis-classified to be part of this globular cluster due to our \HDBSCAN~parameter setting. Namely, the minimum cluster size of 20 prohibited these Sequoia stars from forming their own cluster. Repeating \HDBSCAN~clustering in the full sample space with minimum cluster size reduced to 10, the result indicates Sequoia-like stars are identified separately as their own cluster. We are thus confident that we may safely remove the globular cluster core members and treat the rest of the cluster as a significant overdensity. We can do so given the globular cluster's well-known coordinates ($(l, b) = (277.21,8.63)$ degrees) and angular radius ($0.05$\degree) \citep{harris10} in the sky (see the details of this overdensity in Figure~\ref{Eact_stack_20_merged_summary_no_ngc}). Note that the few stars that are left have no metallicities available. 

There has been a discussion in the literature about the existence of multiple populations within the currently identified retrograde structure named Sequoia, i.e., Arjuna and I'itoi. The three structures are discussed in more detail in \citet{naidu20}. The latter two are discovered in chemical space as separate peaks, with [Fe/H] of $-$1.6, $-$1.2, $<-$2 for Sequoia, Arjuna, and I'itoi, respectively. The lack of metallicity measurements for our cluster prevents us from resolving potential population compositions of this prograde structure. Like GSE, Sequoia is also believed to have most of its stars having a zero mean $v_z$. Thus, our initial cut has preferentially removed Sequoia. Despite that, \HDBSCAN~ identifies an overdensity at high \zmax, signaling that Sequoia may be more kinematically heated and vertically extended than previously thought. 

We associate multiple prograde structures (all in merged cluster \RN{4}, shown in dark red) with the thin and the thick disk, as they have rotational velocity of $\sim-200$~km\,${\text{s}^{-1}}$ and metallicity $\sim -0.5$~dex, consistent with the disk of the Milky Way \citep{bland-hawthorn16}.  
The position of these clusters in the \zmax~vs. eccentricity plane also suggest the possibility of these stars being the high vertical motion end of the velocity distribution of the thin disk and thick disk. We evaluated whether these stars belong to Nyx; crossmatching Nyx and Nyx-2 with the high-stability sample in integrals of motion space yields no match. In fact, the majority of the Nyx stars are cut out by our \zmax\ cut, so we would not expect this cluster to be associated with the Nyx.

\subsection{Comparison with Other Works}
\label{sec:other_group_discussion}

Previous work has been done in identifying local substructures self-consistently with only kinematic data as input. We now compare the results of our clustering algorithm to those of previous studies.

We first discuss the work of \cite{Lovdal22,Ruiz22}, which shares similar scientific goals and approaches as this study. \citet{Lovdal22} adopted a \Gaia~eDR3 sample with crossmatched radial velocities from other spectroscopic surveys than APOGEE and LAMOST. The halo sample, used for clustering, is isolated by applying a $v_{\text{LSR}}$~cut. They opted for the single linkage tree as the clustering algorithm, combined with Mahalanobis distances \citep{mahalanobis36} for final cluster and membership selections. While different, \HDBSCAN~is inherently also a linkage tree algorithm. The clustering space of \citet{Lovdal22} is limited to energy and angular momenta, and uncertainties in these quantities have not been taken into account. 

While both \citet{Lovdal22} and our study identified structures such as the GSE, Helmi Stream, and Sequoia, we do not identify some of the structures (e.g., Thamnos) identified in \citet{Lovdal22} mostly because of the different cuts applied. 
The \zmax~cut has removed the majority of the structures to the point that \HDBSCAN~can no longer pick up any overdensities in phase space. This is likely the case for Thamnos, which is known to be associated with a low stellar mass accretion event that happened a long time ago, a low metallicity, and low total energy. The mean $\langle v_z \rangle$ for Thamnos is consistent with being zero from previous studies, and its current identified member count is lower than that of GSE or Sequoia. Moreover, our average cluster size is significantly smaller due to the different approach we used for selecting robust cluster members with uncertainties considered, along with the different algorithm used. While some clusters identified in their study also appear in similar phase space locations in single realizations of this study, the majority of the stars within these overdensities, as discussed in Section~\ref{sec:uncertainties}, do not robustly belong to said clusters when all random realizations are taken into account. We see in Figure~\ref{vel_stack_Nstack_dist} that most of the stars ever associated with any cluster by \HDBSCAN~in the 100 random realizations have \Nstack~less than 10, i.e., they are flagged as noise 90\% of the time. The robust clusters and their constituents are thus a smaller subset of what one would get by applying \HDBSCAN~to only the full sample with nominal value inputs.

\citet{Ruiz22} used a crossmatched chemical information and mass-age relation to characterize the identified clusters in \citet{Lovdal22}. The membership selection was validated with isochrone fitting and metallicity distribution function study, where there are available matches in spectroscopic surveys. We do not carry out a similar validation because requiring stability of the clusters comes at the expense of statistics. As such, we cannot reliably characterize the progenitors with such few available chemical abundances. While our sample size does allow further study of the structure's color-magnitude diagram and potential comparison with synthetic color-magnitude diagrams, factoring in the observational uncertainties from the data and theoretical uncertainties from the stellar evolution models is not a trivial effort \citep{cignoni10}. We thus defer characterizing our robust clusters with synthetic color-magnitude diagram fitting procedures to a future study. 

We also compare our results with those from \citet{naidu20}, where selections are more based on physical intuition. Again, we recover the GSE, Helmi Stream, and Sequoia, but miss Thamnos for the same reasons mentioned above. We also do not detect any structure that resembles Aleph or Wukong, both reported to have high $J_z$\citep{naidu20}. While Aleph has a relatively high vertical motion, its selection required chemical information which is not available to our study. Wukong is a structure found to be two overdensities near the margin of the GSE in the $E_{tot}~vs.~L_z$ plane. With only kinematic information, we suspect that incorporating uncertainties has caused the cluster to become unstable and thus make it more difficult for \HDBSCAN~to robustly separate it from the GSE and background. We also point out that \citet{naidu20} focused on stars 3 kpc away from the solar neighborhood, so it is also likely that these two structures are only present/or dominate at high galactic latitude, and thus we do not have large enough statistics within our sample to identify such kinds of structures. 

Our clustering results give two clusters associated with the GSE. \citet{donlon22} discussed the GSE being composed of multiple components, namely the VRM, Nereus, and Cronus, using kinematics and metallicities for \Gaia~eDR3 dwarf stars. We compare our pre-merging clusters 1 and 3 in the integrals of motion space with the three radial merger events identified by \citet{donlon22}. Our pre-merging cluster 1 resembles the VRM and Nereus with its high energy and double-lobed $v_r$~velocity profile, but the average metallicity of our cluster [Fe/H]~$\sim -1.07$ is higher than those of the VRM and Nereus at $-$1.7 and $-$2.1 respectively. Similarly, the lower energy cluster 3 has slightly higher metallicity [Fe/H]~$\sim -0.89$ than the Cronus at $-$1.2. Further, while the Cronus was shown to be prograde with $L_{z} \sim -400$~kpc\,km\,${\text{s}^{-1}}$, our low energy GSE-like cluster is consistent with being non-rotating. In short, while our clustering results in the integral of motion space separate the GSE into high-energy and low-energy components, the metallicities do not show as significant a difference as that shown in \citet{donlon22}. The discrepancy may be a combined result of systematic differences in metallicity derivation, given that \citet{donlon22} used photometric metallicity estimates calibrated using a collection of spectroscopic survey catalogs in the literature (see Section 2.3 in \citet{kim22}) while we only used LAMOST DR6, and/or potential contamination from in-situ stars in our clusters given that our clustering is based solely on kinematic properties. For a similar reason, we are not able to distinguish potential constituents in our high eccentricity clusters the same way as in \citet{myeong22}, where chemical abundances from APOGEE DR17 \citep{abdurrouf22} and GALAH DR3 \citep{buder21} are used to construct the chemo-dynamical space for Gaussian Mixture Modeling.

\section{Interpretation and Conclusions}

In this section, we discuss both the challenges in identifying local structures using clustering algorithms as well the potential that new approaches can deliver with current and future data with regards to the detailed exploration of the Galaxy's assembly and accretion history.

In general, while satellite galaxies that have merged or are in the process of merging with the Milky Way can principally be extracted with clustering tools, the observed signal is often too weak compared to the background (e.g. structures within the disk) and identification becomes even more challenging with increasing measurement uncertainties. Accordingly, the current standard practice of largely mitigating unreliable selections has been focused on using either chemical abundance information or isochrone fitting techniques to validate the stellar membership. However, this approach, when applied to the entirety of the Galaxy, will be limiting going forward; chemical information is severely restricted by the difficulty in obtaining medium or high-resolution spectroscopic metallicities and abundances for quantities of stars that are comparable to those of large-scale astrometric and photometric surveys such as \emph{Gaia}. Isochrone fitting of a collection of stars is dependent on the completeness and selection function of any sample. These methods of a posteriori resorting the data also partially undermine the purpose of using an unsupervised machine learning algorithm to handle large data sets and make objective decisions unaffected by human biases. 

Another problem on its own is measurement uncertainties associated with input kinematic data. Unfortunately, the current scope of our study is severely limited by detailed and precise 6D kinematics information, especially for fainter stars that probe deep into the galactic halo, when measurement uncertainties become larger or no information is available at all. Membership assignments by hierarchical clustering algorithms such as \HDBSCAN~ are generally unreliable when uncertainties are not taken into account. For example, when considering stars that are associated with a random cluster in more than one out of 100 random clustering realizations, the majority of the stars are often assigned to be noise in the other 99 realizations. 
However, this is not too surprising. Hierarchical clustering algorithms, e.g., single-linkage algorithms such as \HDBSCAN, rely on first building some form of a hierarchical merger tree, which can alter significantly when data points are shifted within their respective uncertainties. A changed tree will naturally lead to altered findings and as such will yield unreliable results when it comes to identifying new overdensities and their constituents. Thus, the results from applying such algorithms straight to the nominal mean values of the input parameters cannot be trusted.
\textit{We thus stress the need for numerical techniques that produce the most reliable and reproducible clustering results, by taking advantage of incorporating existing observational and astrometric uncertainties in the clustering algorithms. Specifically, we propose a more robust method of utilizing \HDBSCAN~in identifying local stable stellar overdensities in kinematic space.}

To overcome these challenges and to identify robust and reliable clusters, in this study, we use \Gaia eDR3 astrometry data together with line-of-sight radial velocity from \Gaia DR2 to identify clusters in the solar neighborhood using \HDBSCAN.
In our novel procedure, we successfully incorporate measurement uncertainties in the clustering algorithm to produce stable and robust clusters and recover previously known structures in the Milky Way, specifically the GSE, Sequoia, the Helmi Stream, and the globular clusters NGC 3201 and NGC 104. We achieved this by resampling the input data, repeating the clustering process, and treating each random realization equally. We then incorporate measurement uncertainties in the clustering algorithm by resampling the input data, repeating the clustering process, and treating each random realization equally. This technique allows us to confidently isolate the most stable and robust cluster members by stacking the clustering results from each realization and removing those that are associated with noise in more than 20 (out of 100) realizations. These cluster members are then re-clustered into our final set of stable clusters.

When properly incorporating measurement uncertainties in a routine fashion, the resulting robust cluster members will naturally be much fewer than those that would be obtained from applying algorithms to only the nominal values. We have thus focused in this study on retrieving stellar core members of structures near us. Regardless, these results already provide candidate structures that could be targeted for follow-up observations with future spectroscopic surveys, such as 4MOST \citep{4most19} and WEAVE \citep{weave12}. Since obtaining spectroscopic data requires significant amounts of time, as more clusters as well as cluster candidates are being identified, targeted follow-up will be maximally efficient, ensuring a high discovery rate. 

To further increase efficiency and maximal identification of as many cluster member stars as possible, as a next step we anticipate expanding our technique to include a "cluster re-building" component in which we start with the robust clusters and then carry out an additional member selection with progressively loosened stability criteria to eventually re-build the full clusters step-by-step, still with uncertainties factored in. This approach makes use of the fact that the existence of the cluster is no longer in question, and instead focuses efforts on finding the complete stellar membership.

More broadly, our technique of providing precise identifications of substructure signatures has the potential to go beyond finding just accreted structures. In particular, in the galactic disk, studies of moving groups have long revealed many overdensities of stars moving in common orbits \citep{antoja18}, with the most dominant ones being the branches of Sirius, Coma Berenices, Hyades-Pleiades, and Hercules. These local overdensities, proposed to be induced by dynamical resonances, past perturbations, and/or non-axisymmetric features of the Milky Way, are all prime targets of research and provide valuable insights into the formation and evolution of the galactic disk (see, e.g., Appendix B of \citet{trick2019} for a short review). New moving groups are still being discovered thanks to constantly improving methodologies \citep{ramos18,bernet22,lucchini22}. Identifying these overdensities in the disk with high confidence requires careful treatment of the uncertainties, as the structures will most likely be accompanied by strong background noise. 
\cite{lucchini22}, for example, incorporated various measurement uncertainties with wavelet transforms, which is another promising way to maximize unsupervised data analysis outputs. 

These novel data analysis approaches, and as more and deeper data becomes available, from both space missions, such as \textit{Gaia}, and ground missions, such as the Vera Rubin Observatory Legacy Survey of Space and Time (LSST) \citep{2019ApJ...873..111I}, showcase the huge discovery potential for stellar (sub)structures in the Milky Way. We thus anticipate that many more delicate, thus far hidden clusters will be discovered. But success will rest solely on our ability to rely on unsupervised machine learning algorithms, such as \HDBSCAN, to provide trustworthy outputs. We can only trust algorithms when we factor in measurement uncertainties that, of course, are always present in observational astronomy. 

In closing, understanding in detail how accurate these techniques and algorithms are given the current level of uncertainties in observational measurements is an important step towards building a complete picture of the Milky Way, as well as towards providing improved modeling capabilities of the local DM phase space distribution. 


We thus encourage future clustering studies to begin testing and incorporating uncertainties into their new generation of techniques and unsupervised machine learning algorithms, to acquire reliable and robust selections of substructures and their member stars based on kinematic information only.

\begin{acknowledgements}

We thank M.~Lisanti, T.~Nguyen, C.~Roche, and N.~Shipp for helpful discussions.
X. O. and A.F. acknowledge support from NSF grant AST-1716251.

This work presents results from the European Space Agency (ESA) space mission \emph{Gaia}. \Gaia data are being processed by the \Gaia Data Processing and Analysis Consortium (DPAC). Funding for the DPAC is provided by national institutions, in particular the institutions participating in the Gaia MultiLateral Agreement (MLA). The Gaia mission website is \url{https://www.cosmos.esa.int/gaia}. The Gaia archive website is \url{https://archives.esac.esa.int/gaia}.

This work used Stampede-2 under allocation number TG-PHY210118, part of the Extreme Science and Engineering Discovery Environment (XSEDE), which is supported by National Science Foundation grant number ACI-1548562.

This research has made use of NASA's Astrophysics Data System Bibliographic Services; the arXiv pre-print server operated by Cornell University; the SIMBAD and VizieR databases hosted by the Strasbourg Astronomical Data Center

\end{acknowledgements}

\software{%
matplotlib \citep{hunter07},
numpy \citep{vanderwalt11},
scipy \citep{jones01},
astropy \citep{2013A&A...558A..33A,2018AJ....156..123A},
gala \citep{gala_software}, and 
AGAMA \citep{vasiliev19}.}

\clearpage

\bibliographystyle{aasjournal}
\bibliography{xou}

\setcounter{equation}{0}
\setcounter{figure}{0}
\setcounter{table}{0}
\setcounter{section}{0}
\makeatletter
\renewcommand{\theequation}{S\arabic{equation}}
\renewcommand{\thefigure}{S\arabic{figure}}
\renewcommand{\thetable}{S\arabic{table}}

\appendix 

\begin{figure}[h]
\begin{center}
\includegraphics[angle=0,width=0.45\textwidth]{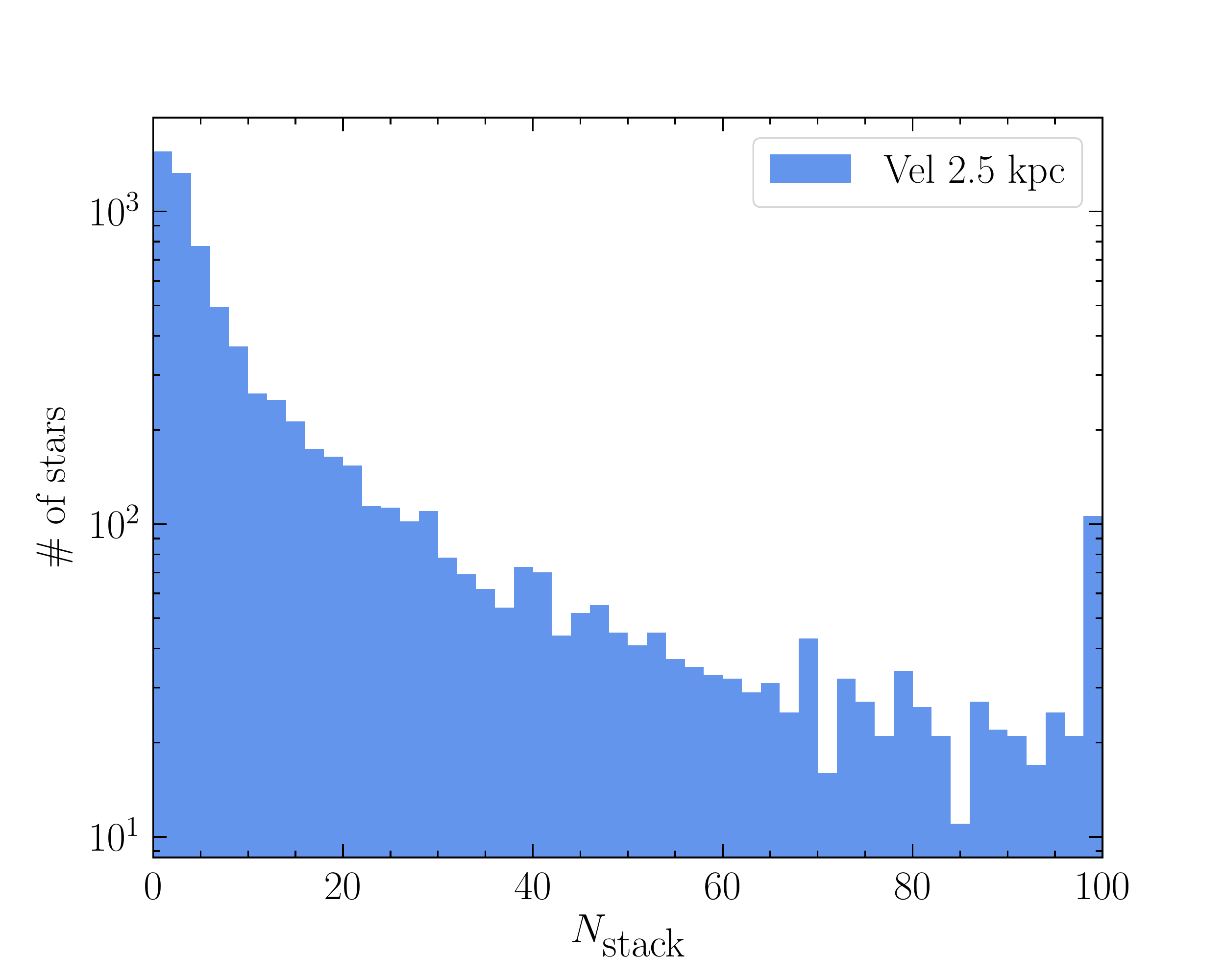}
\end{center}
\caption{Distribution of \Nstack~for the stacked full subsample with \zmax~2\,$\sigma$ above 2.5 kpc from the cylindrical velocity space.}
\label{vel_stack_Nstack_dist}
\end{figure}
\begin{figure}[h]
\begin{center}
\includegraphics[angle=0,width=0.3\textwidth]{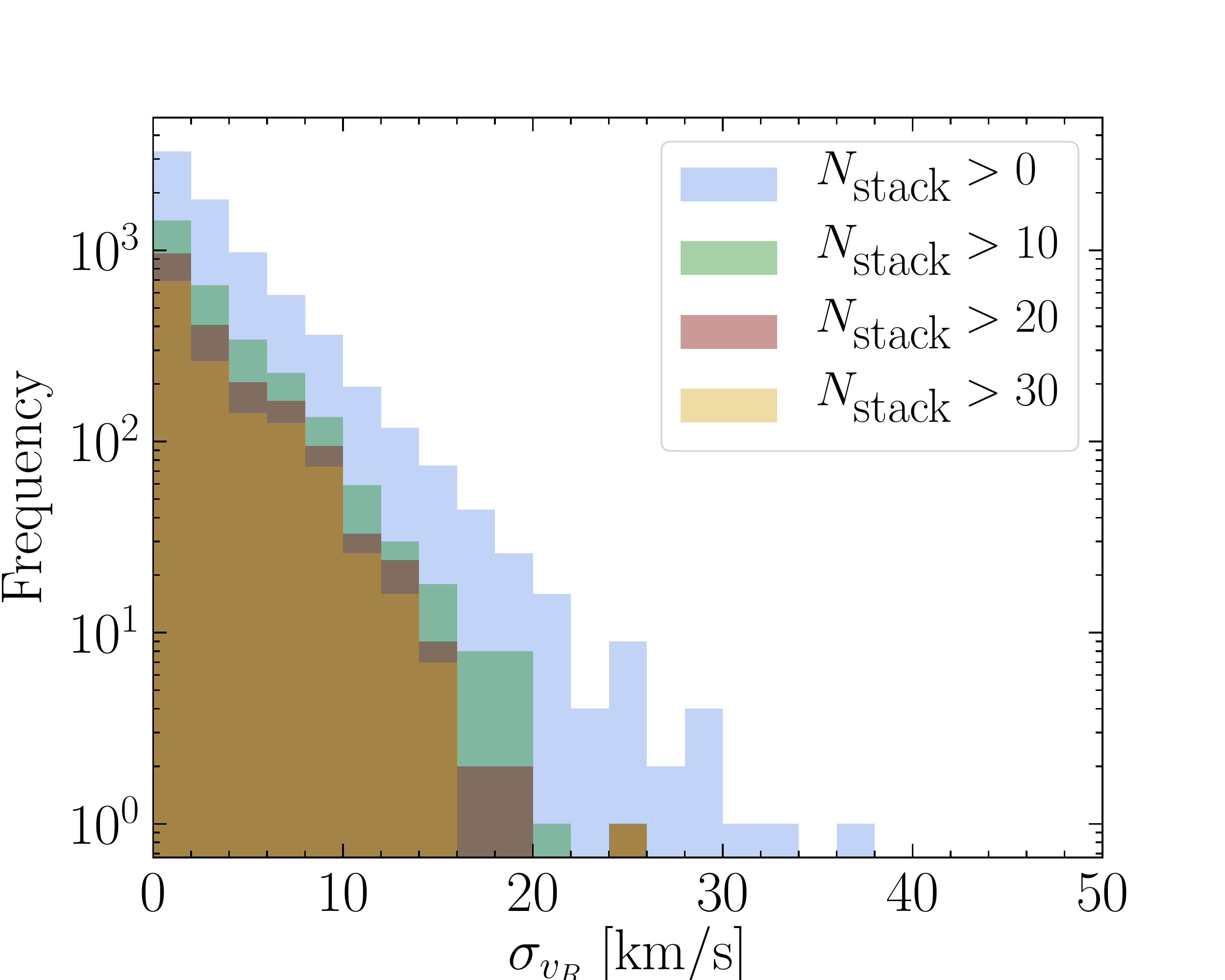}
\includegraphics[angle=0,width=0.3\textwidth]{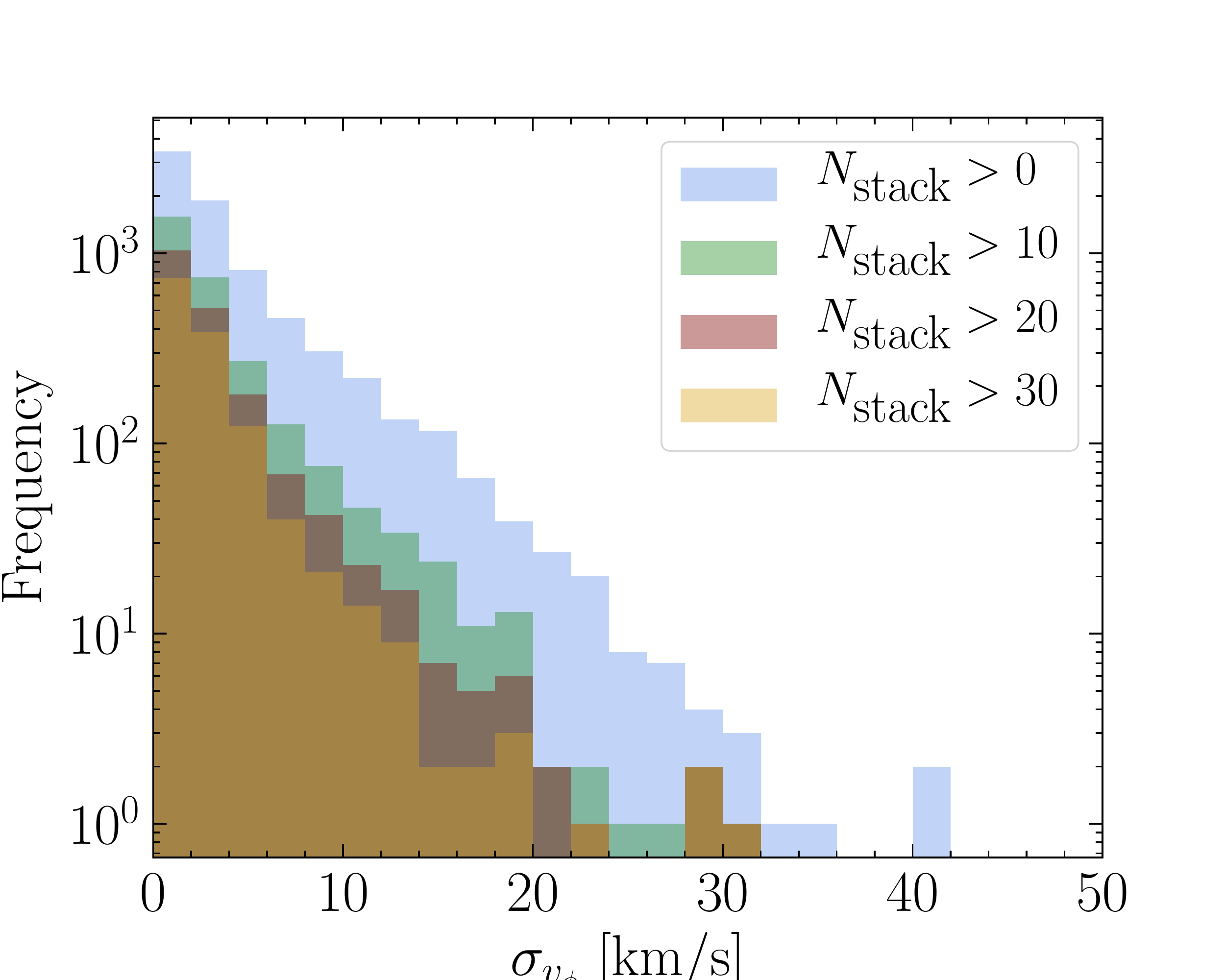}
\includegraphics[angle=0,width=0.3\textwidth]{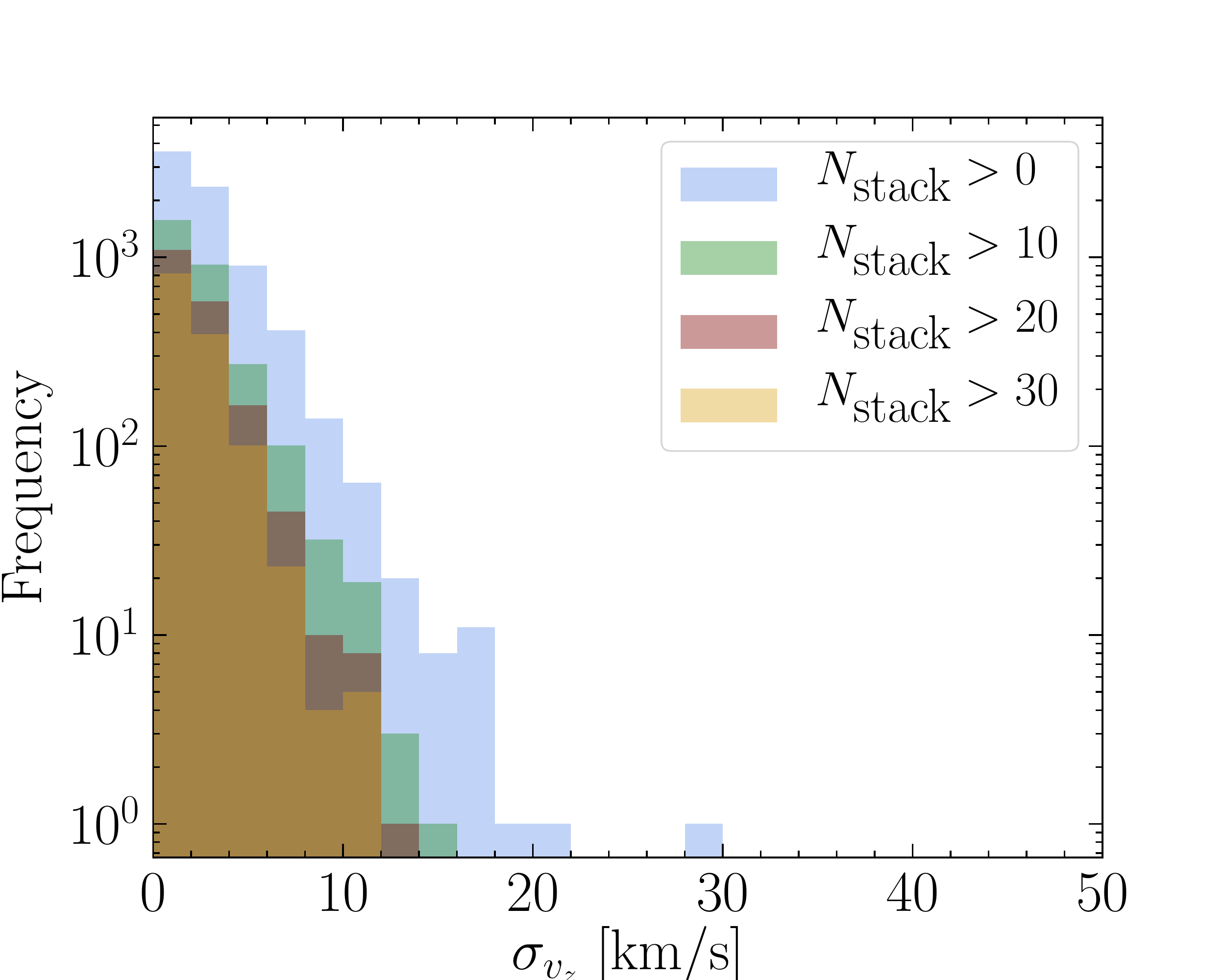} \\
\includegraphics[angle=0,width=0.3\textwidth]{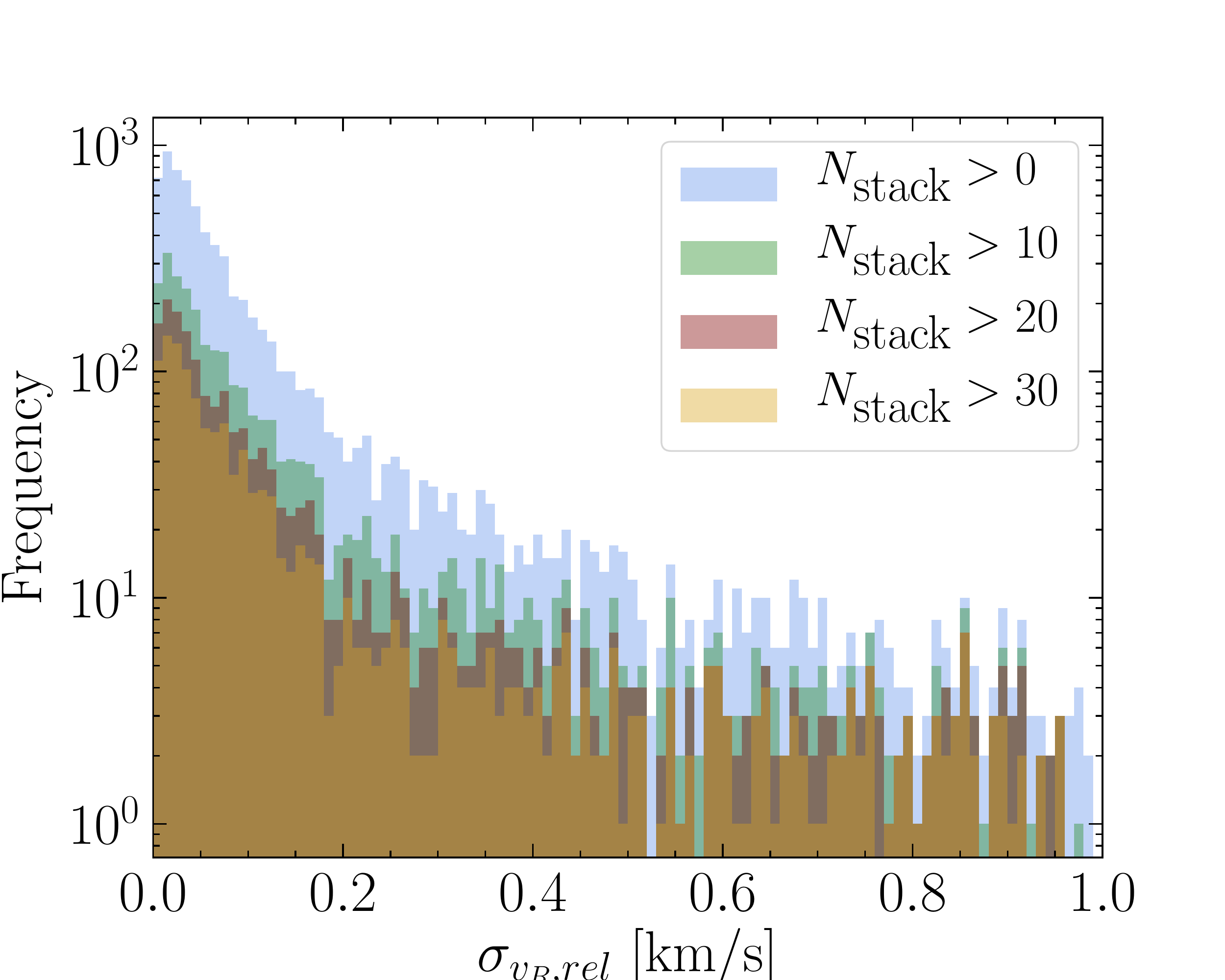}
\includegraphics[angle=0,width=0.3\textwidth]{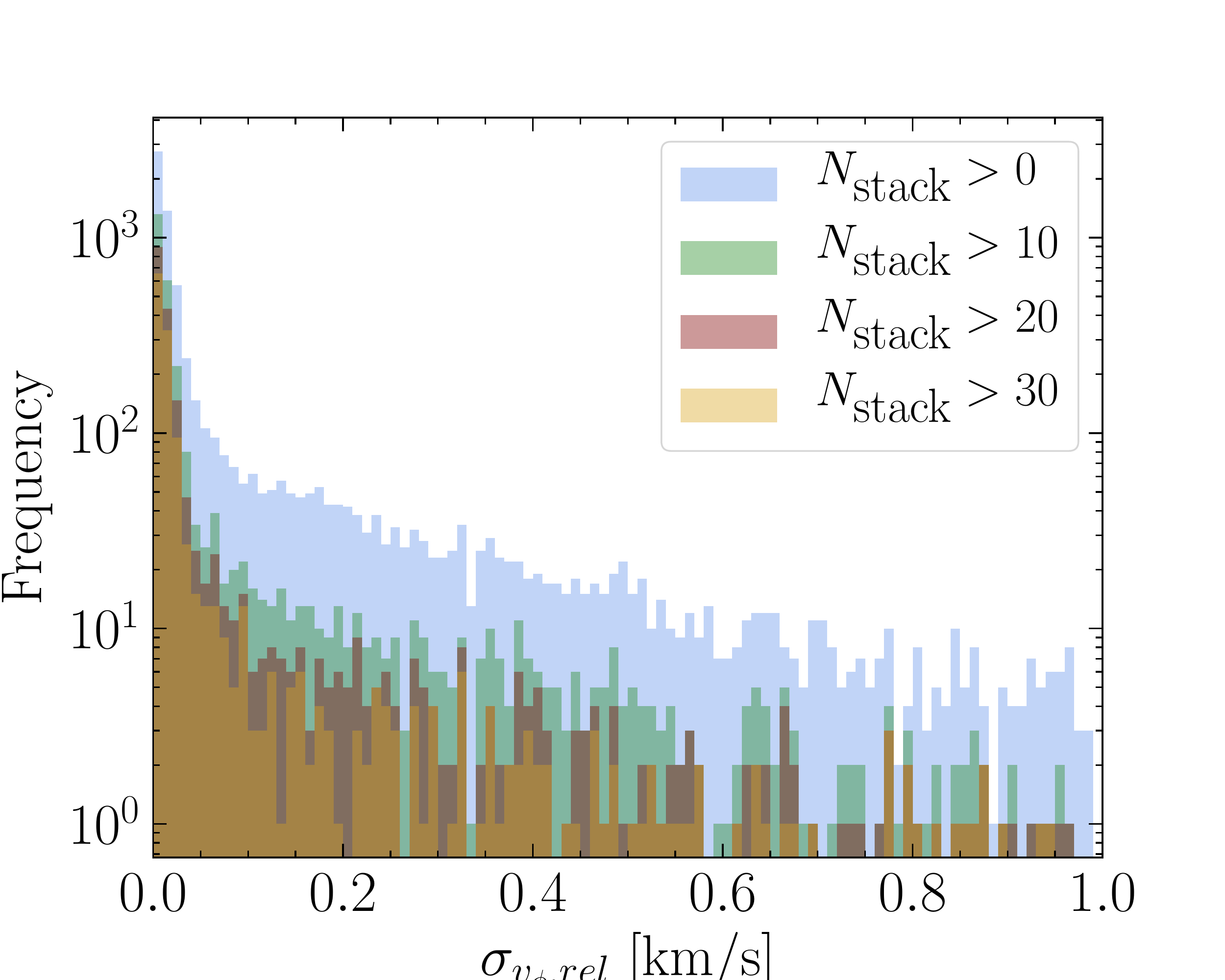}
\includegraphics[angle=0,width=0.3\textwidth]{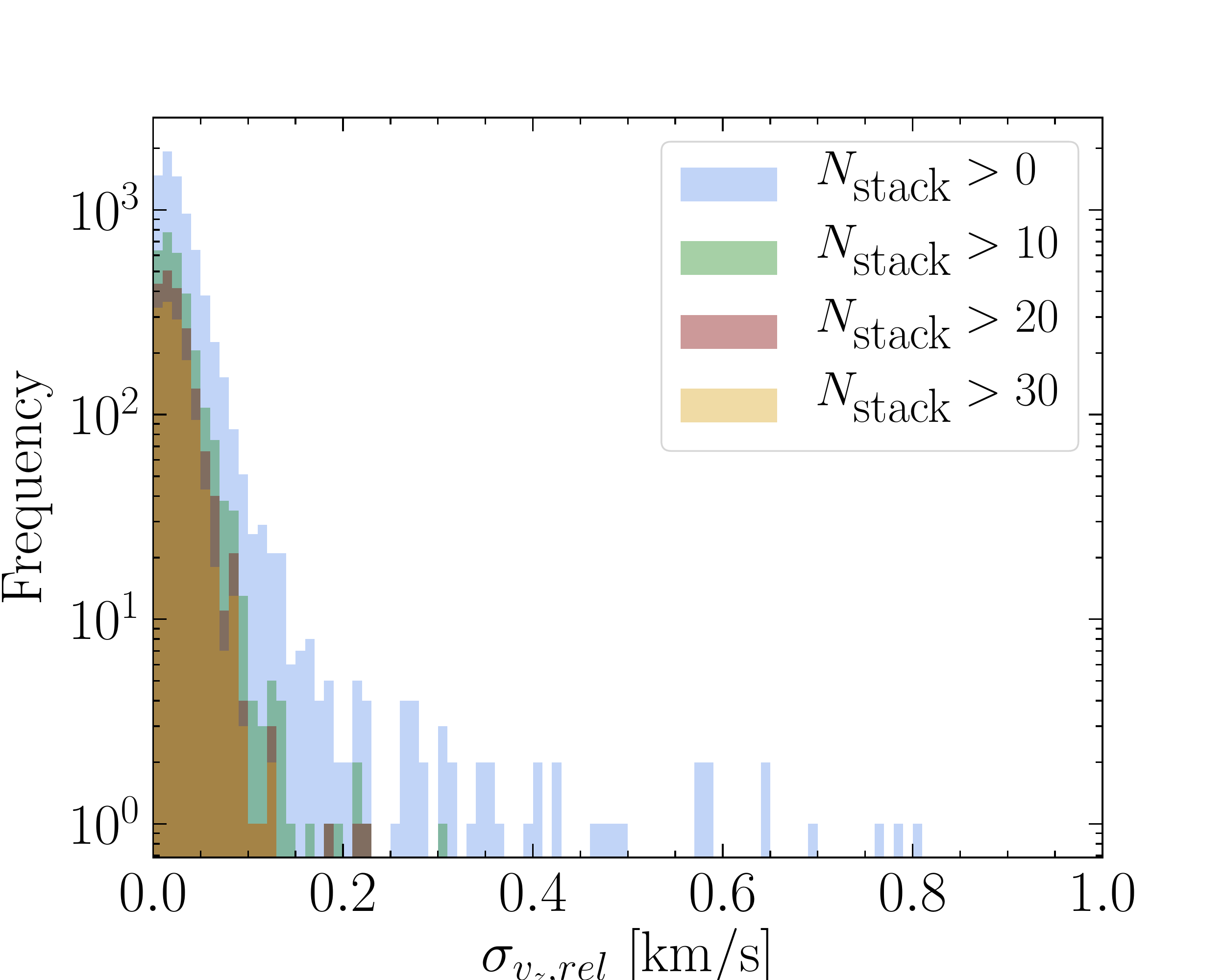}
\end{center}
\caption{Absolute and relative error distribution of cylindrical velocities for the stacked full subsample with \zmax~2\,$\sigma$ above 2.5 kpc at different \Nstack~cuts. Note that the y axis is on logarithmic scale.}
\label{vel_stack_err_dist}
\end{figure}
\begin{figure}[h]
\begin{center}
\includegraphics[angle=0,width=6.7in]{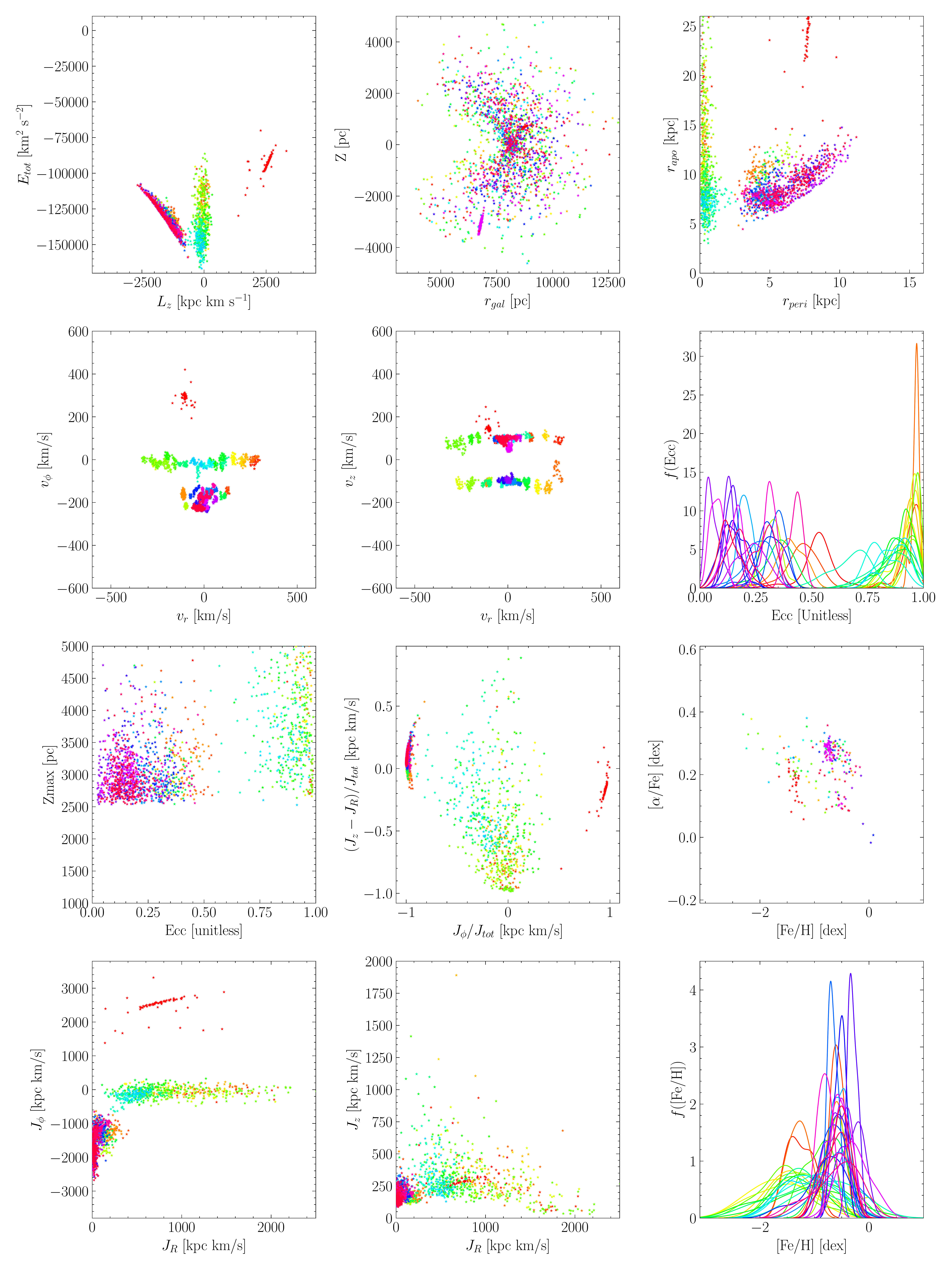}
\end{center}
\caption{Summary plot for clusters that \HDBSCAN~identified from stacked full subsample with \zmax~2\,$\sigma$ above 2.5 kpc from the cylindrical velocity space.}
\label{vel_stack_full_summary}
\end{figure}
\begin{figure*}[h]
\begin{center}
\includegraphics[angle=0,width=6.7in]{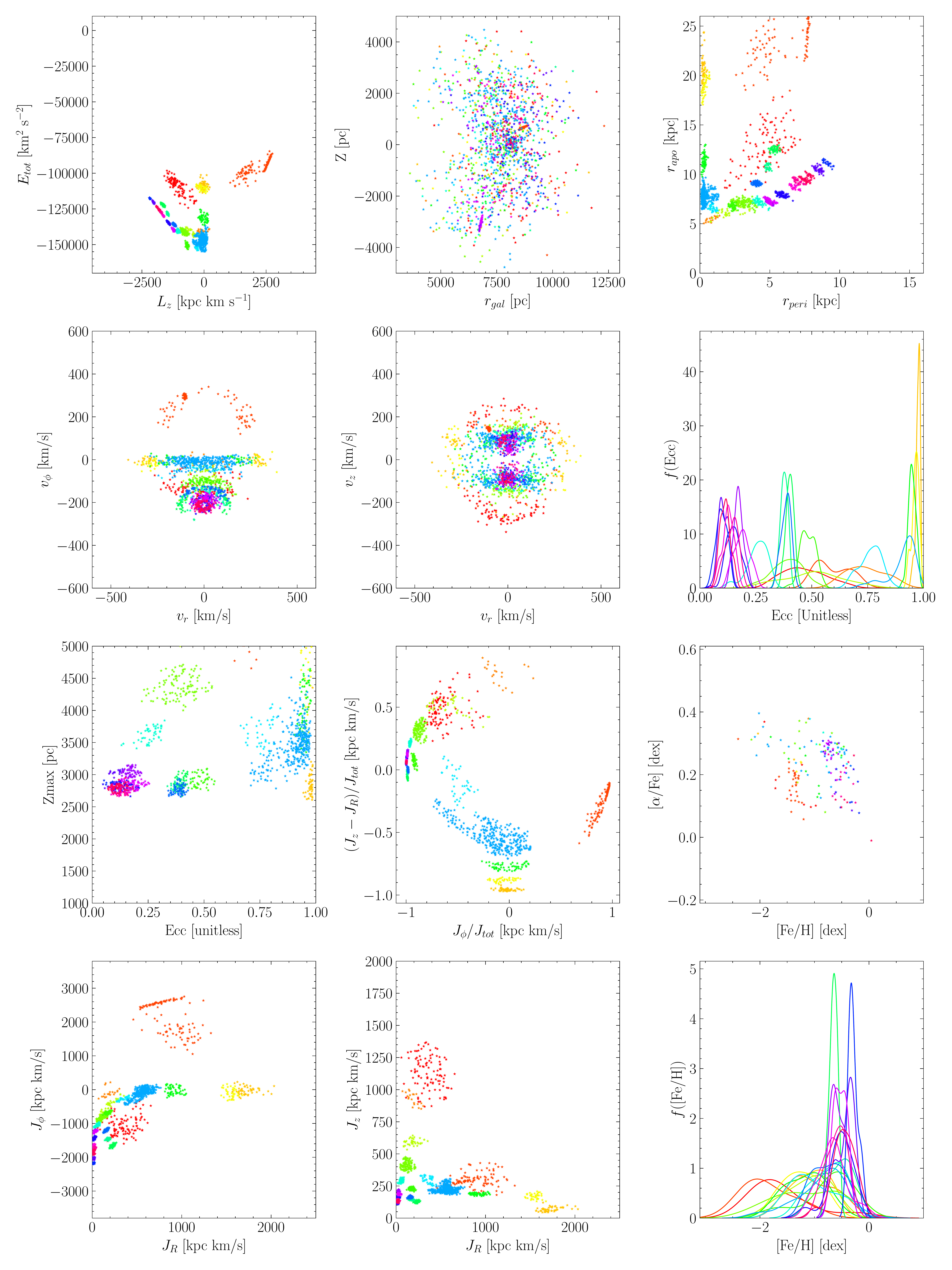}
\end{center}
\caption{Summary plot for clusters that \HDBSCAN~identified from stacked full subsample with \zmax~2\,$\sigma$ above 2.5 kpc from the Energy + 3D action space.}
\label{Eact_stack_full_summary}
\end{figure*}
\begin{figure*}
\begin{center}
\includegraphics[angle=0,width=0.90\textwidth]{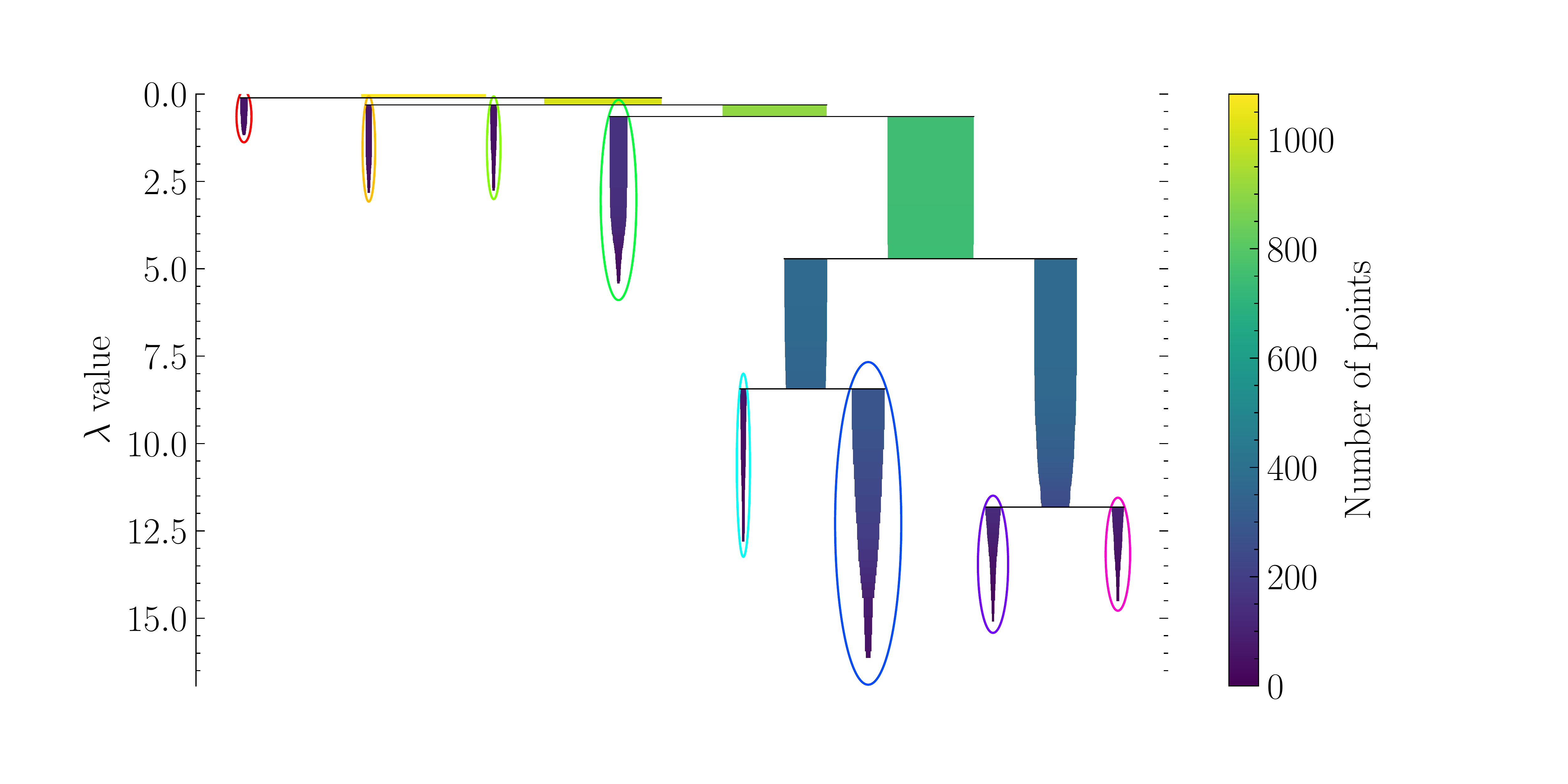}
\end{center}
\caption{Condensed merger tree for the stacked subsample with \Nstack$>20$ and \zmax~2\,$\sigma$ above 2.5 kpc from the Energy + 3D action space.}
\label{Eact_stacked_20_2p5_tree}
\end{figure*}
\begin{figure*}
\begin{center}
\includegraphics[angle=0,width=6.7in]{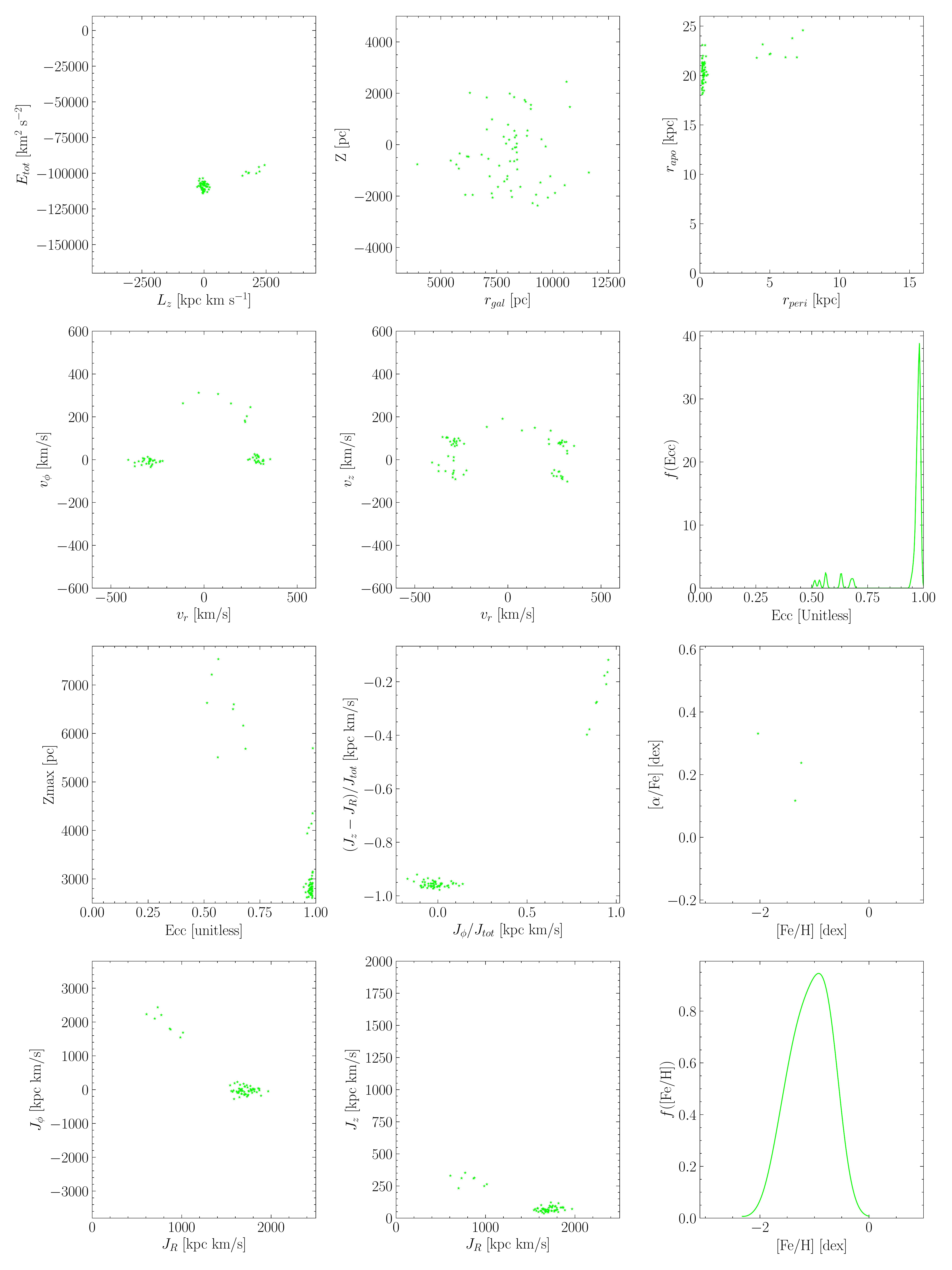}
\end{center}
\caption{Summary plot for cluster 1 from \HDBSCAN~identified from stacked stable (\Nstack$>20$) subsample with \zmax~2\,$\sigma$ above 2.5 kpc from the Energy+Action space with NGC 3201 removed.}
\label{Eact_stack_20_merged_summary_no_ngc}
\end{figure*}

In this appendix, we show the figures associated with the \Nstack~cuts. Fig~\ref{vel_stack_Nstack_dist} shows the number of stars associated with any cluster at \Nstack~number of realizations. Figure~\ref{vel_stack_err_dist} shows the error distribution of stars with various \Nstack~cuts. We also include Figure~\ref{vel_stack_full_summary} and \ref{Eact_stack_full_summary}, showing the clustering result on the velocity and integrals of motion spaces respectively, without placing any \Nstack~cut on the stacked sample. As expected they include a much larger number of clusters. We also include Figure~\ref{Eact_stacked_20_2p5_tree}, the equivalent of Figure~\ref{vel_stacked_20_2p5_tree} in integrals of motion space. Finally, Figure~\ref{Eact_stack_20_merged_summary_no_ngc} shows the resulting clustering, having removed cluster NGC~3201 (see related discussion in Section~\ref{sec:integrals}).

\end{CJK*}
\end{document}